\documentclass[a4paper, 12pt]{article} \textheight8.2in
\usepackage{textcomp}
\usepackage{amsmath}
\usepackage{amssymb}
\usepackage{amsthm}
\usepackage{setspace}
\usepackage{mleftright}
\usepackage{multirow}
\usepackage[utf8]{inputenc}
\usepackage{filecontents}
\usepackage{graphicx}
\usepackage{sectsty}
\usepackage{titletoc}
\usepackage[toc,page]{appendix}
\graphicspath{ {figures/} }
\usepackage{array}
\usepackage{float}
\usepackage{scalerel,stackengine}
\usepackage{enumitem} 
\usepackage{mathtools}
\usepackage{color, graphics, graphicx}
\usepackage{xcolor,colortbl}

\usepackage{floatpag}

\newtheorem{theorem}{Theorem}
\numberwithin{equation}{section}
\numberwithin{theorem}{section}
\newcommand{\mbf}[1]{\mbox{\boldmath $#1$}}
\newcommand{\bth}{{\mbf \theta}}

\def\s{\sum_{t=1}^n}
\def\d{\dot}
\def\h{\hat}

\def\nn{\nonumber}
\def\bth{\mbox{\boldmath$\theta$}}

\def\bTheta{\mbox{\boldmath$\Theta$}}

\def\vs{\vskip}

\def\R{{\mathbb R}}

\def\G{{\mbf G}}

\def\0{{\mbf 0}}

\title{M-Estimation in GARCH Models\\ in the Absence of Higher-Order Moments} 
\date{}  
\author{
Marc Hallin\\
{\small ECARES and D\'epartement de Math\'ematique}\\
{\small Universit\'e Libre de Bruxelles CP 114/4}\\
{\small Avenue F.D.\  Roosevelt 50,  
B-1050 Bruxelles Belgium} \\ 
{\small E-mail: mhallin@ulb.ac.be}\\
\, \\  
Hang Liu\\
{\small International Institute of Finance, School of Management}\\
 {\small University of Science and Technology of China}\\
{\small Hefei, 
 Anhui  
230026 China} \\ 
{\small E-mail: hliu01@ustc.edu.cn}
\\ 
\, 
\\  and  \\ \, \\ 
Kanchan Mukherjee\\ 
{\small Department of Mathematics and Statistics}\\ 
{\small Lancaster University, United Kingdom
LA1 4YF}\\ 
{\small E-mail: k.mukherjee@lancaster.ac.uk}
}

\begin{document}
\maketitle
\begin{quote}
\begin{abstract}  We consider a class of M-estimators of the parameters of a GARCH$(p,q)$ model. 
These estimators involve score functions and, for adequate choices of the score functions, 
are asymptotically normal under milder moment assumptions than  the usual quasi maximum likelihood, which makes them more reliable in the presence of heavy tails.  We also consider  weighted bootstrap approximations of the distributions of these M-estimators and establish their validity. Through extensive simulations, we demonstrate the robustness of these M-estimators under heavy tails and conduct a comparative study of the performance (bias and mean squared errors) of various score functions  and  the accuracy (confidence interval coverage rates) of their bootstrap approximations. In addition to the GARCH~(1, 1) model, our simulations also involve higher-order models such as GARCH~(2, 1) and GARCH~(1,~\!2) which  so far have  received relatively little attention in the literature. We also consider the case of order-misspecified   models. 
Finally, we use our M-estimators in the analysis of two real financial time series fitted with GARCH~(1, 1) or GARCH~(2, 1) models. 

%
\end{abstract}
\vs .2cm


Short title: {\it M-estimation in GARCH models.}\\
\end{quote}

\pagenumbering{arabic} \pagestyle{plain} \setcounter{page}{0}
\addtocounter{page}{2}
\section{Introduction}
\setcounter{equation}{0} 

Generalized Auto Regressive Conditional Heteroscedastic (GARCH) models have been used extensively to analyze the volatility or the instantaneous variability in financial time series. This is a domain in which Professor Masanobu Taniguchi and his coauthors have a number of impactful papers (Lee and Taniguchi, 2005; Taniai et al., 2012) and two influential  monographs (Taniguchi et al., 2008 and 2014).

A stochastic process ${\cal X}\coloneqq \{X_t; t \in \mathbb{Z}\}$  is said to follow a GARCH~$(p, q)$ model if
\begin{equation}\label{m1}
X_t= \sigma_{t}\epsilon_t,\quad t\in\mathbb{Z}
\end{equation}
where $\{\epsilon_t; t \in \mathbb{Z}\}$ are unobservable i.i.d.\ errors with symmetric distribution around zero and~$\{\sigma_t; t \in \mathbb{Z}\}$ is a solution of 
\begin{equation}\label{m2}
\sigma_{t}= (\omega_0+\sum_{i=1}^p \alpha_{0i}X^2_{t-i}+\sum_{j=1}^q \beta_{0j} \sigma^2_{t-j})^{1/2}, \quad\,t \in \mathbb{Z},
\end{equation}
for some  $\omega_0>0$, $\alpha_{0i}>0$, $i=1,\ldots,p$, and  $\beta_{0j}> 0$, $ j=1,\ldots,q$. Mukherjee (2008) introduced a class of M-estimators for estimating the GARCH parameter 
\begin{equation}\label{thetanot}
\bth_0\coloneqq (\omega_0, \alpha_{01}, \ldots, \alpha_{0p}, \beta_{01}, \ldots, \beta_{0q})^{\prime}
\end{equation}
based on an observed finite realization  $\{X_t; 1 \le t \le n\}$ of $\cal X$. Depending on the choice of a score function, these M-estimators are asymptotically normal under milder moment assumptions on the error distribution than the commonly-used quasi maximum likelihood estimator (QMLE). Mukherjee (2020) further considered a class of weighted bootstrap methods to approximate the distributions of these estimators and established their asymptotic validity. In this paper, we discuss an iteratively re-weighted algorithm to compute these M-estimates and the corresponding bootstrap estimates with emphasis on the so-called Huber, $\mu$-, and Cauchy M-estimates, which so far were not given much attention  in the literature. This iteratively re-weighted algorithm turns out to be particularly useful in the computation of  bootstrap replicates since it avoids re-evaluating some core quantities for each new bootstrap sample. 

The class of M-estimators of Mukherjee (2008) includes the (Gaussian) QMLE as a special case. The asymptotic normality of the QMLE and the asymptotic validity of bootstrapping it are well-known  classical results which, however,  require the assumption of finite fourth-order moment of the error distribution. The same class also contains other less-known M-estimators, such as the $\mu$-estimator and Cauchy-estimator,  which are asymptotically normal under milder moment assumptions and hence 
should be considered as attractive alternatives to the QMLE. One of the objectives of this paper is to study the performance of these estimators through simulations and use them for the empirical study on some interesting datasets.    

In an earlier work, Muler and Yohai (2008) analyzed the Electric Fuel Corporation (EFCX) time series and fitted a GARCH~(1, 1) model. Using exploratory analysis, they detected the presence of outliers and considered estimation of the GARCH parameters based on various robust methods. It turns out that the estimates based on different methods vary widely and this makes their study somewhat inconclusive as to which robust methods should be preferred in similar situations. In this paper, we show how M-estimates can be used for making such choice. 

In a different direction, Francq and Zako{\"\i}an (2009) stressed the importance of considering higher-order GARCH models such as the GARCH~(2, 1) in the context of analyzing financial data.  Computational results and simulation studies for such models, however,  are rather scarce in the literature.  Our simulation study and empirical applications therefore include  higher-order models such as GARCH~(2, 1)  and  GARCH~(1, 2).

The main contributions of the paper are as follows. We implement a very general algorithm for computing a variety of M-estimators and demonstrate their importance in the analysis of real data. We consider situations when the error distributions are possibly heavy-tailed or when a higher-order GARCH model is needed for fitting the data. We provide results and analysis of extensive simulation study based on M-estimators which are asymptotically normal under weak moment assumptions on error distribution. Finally, we study the effectiveness of the bootstrap approximation of the distribution of M-estimators.

The paper is organized as follows. Sections \ref{sec.Mest} and \ref{sec.BootMest} set the background. 
In particular, Section \ref{sec.Mest} considers the class of M-estimators and provides several examples. Section~\ref{sec.BootMest} contains the bootstrap formulation and its  asymptotic validity. Section~\ref{sec.comMest} discusses some of the computational aspects of M-estimators  and their bootstrapped versions. Section~\ref{sec.simMest} reports simulation results for various M-estimators. Section \ref{sec.simBootMest} compares the bootstrap approximations of  M-estimators with the classical asymptotic normal  approximation. Section~\ref{sec.MestData} analyzes two real financial time series data. 
  
\section{M-estimation of  GARCH parameters}\label{sec.Mest}
\subsection{A class of M-estimators}
Throughout,  we write $\d{g}$ for the derivative and $\d{{\bf g}}$ for the gradient of  a differentiable  function~$g$, $\mbox{sign}(x)$ for $I(x>0)-I(x<0)$, and $\log^+(x)$ for~$I(x>1)\log(x)$ when $x>0$. Also, $\epsilon$ represents a generic random variable with the same distribution as the errors $\{\epsilon_t\}$ in (\ref{m1}).  

Consider $H(x)\coloneqq x\psi(x)$, $x \in \R$ where $\psi: \R \rightarrow \R$ is an odd and differentiable function at all but possibly a finite number of points;  
denote by~${\cal D} \subseteq \R$  the set of points where $\psi$ is differentiable and by~$\bar{\cal D}$ its complement. Since  $\psi$ is an odd function, $H$ is an even function.  
Functions $H$ of this type will be used as {\it score functions} in the M-estimation procedures described below. Examples are as follows. \medskip

\indent {\bf Example 1.} QMLE score function:  $\psi(x)=x$ ($\bar{\cal D}=\phi$, the empty set), $H(x)=x^2$.\smallskip 

\indent {\bf Example 2.} LAD score function:  $\psi(x)=\mbox{sign}\,(x)$ ($\bar{\cal D}=\{0\}$), $H(x)=|x|$. \smallskip 

\indent {\bf Example 3.} Huber's $k$ score function:  $\psi(x)=xI(|x|\le k)+ k\,\mbox{sign}\,(x)I(|x|>k)$,
where~$k>0$  is a known constant  ($\bar{\cal D}=\{-k, k\}$), $H(x)=x^2I(|x|\le k)+ k |x|I(|x|>k)$.\smallskip 
  
\indent {\bf Example 4.}  Maximum likelihood   (MLE) score function:  $\psi(x)=-\d{f}(x)/f(x)$, where $f$ is the actual density of $\epsilon$, assumed to be known,  and $H(x)=x\{-\d{f}(x)/f(x)\}$.\smallskip 

\indent {\bf Example 5.} $\mu$ score function: 
$\psi(x)=\mu \,\mbox{sign}(x)/(1+ |x|)$, where $\mu>1$ is a known constant ($\bar{\cal D}=\{0\}$), $H(x)=\mu |x|/(1+ |x|)$ (a bounded score function).\smallskip 

\indent {\bf Example 6.} Cauchy score function: $\psi(x)=2x/(1+x^2)$, $H(x)=2x^2/(1+x^2)$  (a bounded score function).\smallskip 

\indent {\bf Example 7.} Exponential pseudo-maximum likelihood   score function:   $\psi(x)=\delta_1 |x|^{\delta_2-1} \mbox{sign}(x)$, where $\delta_1>0$ and $1< \delta_2 \le 2$ are known constants ($\bar{\cal D}=\{0\}$),  
$H(x)=\delta_1 |x|^{\delta_2}$.\smallskip

Assume that for some $\kappa_1 \ge 2$ and $\kappa_2>0$, 
\begin{equation}\label{rep**}
{\mathrm E}[|\epsilon|^{\kappa_1}]<\infty\,\,\,\mbox{and}\,\,\, \lim_{t\to 0} {\rm P}[\epsilon^2 <t]/ t^{\kappa_2}=0.
\end{equation}
Then $\sigma^2_t$ from (\ref{m2}) admits the  unique almost sure representation
\begin{equation}\label{c0}
\sigma^2_t=c_0+\sum_{j=1}^{\infty} c_j X^2_{t-j},\,\,\, t \in \mathbb{Z}
\end{equation}
where $\{c_j; j \ge 0\}$ are defined in (2.9)-(2.16) of Berkes et al. (2003). 
Let $\bTheta$ be a compact subset of $(0, \infty)^{1+p} \times (0, 1)^{q}$. A typical element in $\bTheta$ is denoted by 
$\bth=(\omega, \alpha_{1}, \ldots, \alpha_{p}, \beta_{1}, \ldots, \beta_{q})^{\prime}$. Define the {\it variance function}   $v_t: \bTheta\to\mathbb{R}^+$ by
\begin{equation}\label{cj}
v_t(\bth)=c_0(\bth)+\sum_{j=1}^{\infty} c_j(\bth) X_{t-j}^2, \,\,\,\,\, 
\bth \in \bTheta,\ t \in \mathbb{Z}, 
\end{equation}
where the coefficients $\{c_j (\bth); j \ge 0\}$ are such that, for $\bth=\bth_0$, 
\begin{equation}\label{cjj}
c_j (\bth_0)=c_j, \,\,   j \ge 0
\end{equation}
(Berkes et al. (2003), Section 3 and display~(3.1)).  
Hence the variance function satisfies~$v_{t}(\bth_0)=\sigma_t^2$, $t \in \mathbb{Z}$ and  
 (\ref{m1}) can be rewritten as 
\begin{equation}\label{mmm}
X_t= \{v_{t}(\bth_0)\}^{1/2}\epsilon_t, \;\; 1 \le t \le n.
\end{equation}

Let $H$ denote a score function. The M-estimators are defined as the solutions $\hat{\bth}_n$ 
of~$\widehat{\mbf M}_{n, H}(\bth) = \0$, where
\begin{equation}\label{defn1}
\widehat{\mbf M}_{n, H}(\bth) \coloneqq  \s \bigg\{1-H\{X_t/\h{v}_t^{1/2}(\bth)\}\bigg\}\{\d{\h{\mbf v}}_t(\bth)/\h{v}_t(\bth)\}
\end{equation}
and
\begin{equation}\label{vt.hat.cj}
\hat{v}_t(\bth)\coloneqq c_0(\bth)+I(2 \le t) \sum_{j=1}^{t-1} c_j (\bth) X^2_{t-j}, \quad  \bth \in \bTheta, \,\, 1 \le t \le n
\end{equation}
is the observable approximation of  the variance function $v_t(\bth)$ defined in~(\ref{cj}).  
%
%

The recursive nature of the coefficients $\{c_j(\bth)\}$ greatly simplifies the computation of \linebreak M-estimators, as  discussed in Section \ref{sec.comMest}. For $p,q = 1$ or $2$, these coefficients, for instance,  satisfy the following recursions.\medskip

{\bf Example 1.}   GARCH $(1, 1)$ model: with $\bth=(\omega, \alpha, \beta)^{\prime}$,
$$
c_0(\omega, \alpha, \beta)=\omega/(1-\beta), \  \,c_j(\omega, \alpha, \beta)=\alpha \beta^{j-1},\,\,j \ge 1.
$$

{\bf Example 2.}  GARCH $(2, 1)$ model: with $\bth=(\omega, \alpha_1, \alpha_2, \beta)^{\prime}$,
$$
c_0(\bth)=\omega/(1-\beta), \ \ c_1(\bth)=\alpha_1, \ \ c_2(\bth)=\alpha_2+\beta c_1(\bth)=\alpha_2+\beta \alpha_1, 
$$
and
$$
c_j(\bth)=\beta c_{j-1}(\bth),\quad j \ge 3.\smallskip
$$

{\bf Example 3.}  GARCH $(1, 2)$ model: with $\bth=(\omega, \alpha, \beta_1, \beta_2)^{\prime}$,
$$
c_0(\bth) = \omega/(1-\beta_1-\beta_2),\ \  c_1(\bth) = \alpha, \ \ c_2(\bth) =\beta_1 c_1(\bth)= \beta_1 \alpha,
$$
and 
$$
c_j(\bth) = \beta_1 c_{j-1}(\bth) + \beta_2 c_{j-2}(\bth), \quad j \ge 3.\smallskip
$$

{\bf Example 4.}  GARCH $(2, 2)$ model: with $\bth=(\omega, \alpha_1, \alpha_2, \beta_1, \beta_2)^{\prime}$, 
$$
c_0(\bth) = \omega/(1-\beta_1-\beta_2), \ \ c_1(\bth) = \alpha_1, \ \ c_2(\bth) = \alpha_2+\beta_1\alpha_1,
$$ 
and 
$$
c_j(\bth) = \beta_1 c_{j-1}(\bth) + \beta_2 c_{j-2}(\bth), \quad j \ge 3.
$$
\subsection{Asymptotic distribution of M-estimators }\label{sec.asyMest}

The asymptotic distribution of  M-estimators   is derived under the following assumptions. \medskip 

{\bf Assumptions (A)}   {\it (Model assumptions)}. The parameter space $\bTheta$ is  compact  and $\bth_0$   defined in \eqref{thetanot} belongs to its interior; (\ref{rep**}), (\ref{cj}), and (\ref{mmm}) hold;  $\{X_t\}$ is stationary and ergodic. 
 
{\bf Assumptions (B)}   {\it (On score function)}. 

{\bf (B1)} Associated with the score function $H$, there exists a unique number $c_{H}>0$ such that 
\begin{equation}\label{H1}
{\mathrm E}[H(\epsilon/c_H^{1/2})]=1,\ \ 
{\mathrm E}[H(\epsilon/c_H^{1/2})]^2 < \infty,\,\,\,\mbox{and}\,\,\, 
0<{\mathrm E}\{(\epsilon/c_H^{1/2}) \d{H}(\epsilon/c_H^{1/2})\}< \infty;
\end{equation}
and the transformed parameter 
\begin{equation}\label{thetanoth}
\bth_{0H}\coloneqq (c_H\omega_0, c_H\alpha_{01}, \ldots, c_H\alpha_{0p}, \beta_{01}, \ldots, \beta_{0q})^{\prime}
\end{equation}
is in the interior of $\bTheta$. \medskip

\color{black}
{\bf (B2)} The smoothness conditions:\footnote{These conditions are trivially satisfied by all the examples of score functions $H$ considered above.}
\begin{itemize}
\item[(i)] There exists function $L$ satisfying
\begin{equation*}
|H(es)-H(e)| \le L(e)|s^2-1|, \, e \in \R^1, \, s>0, 
\end{equation*}
where ${\rm E}\log^+\{L(\epsilon/c_H^{1/2})\}< \infty.$
\item[(ii)] There exists function $\Lambda$ such that for $e \in \R^1, \, s>0$,
$es, e \in \cal D$, 
\begin{equation*}
|\d{H}(es)-\d{H}(e)| \le \Lambda(e)|s-1|, 
\end{equation*}
where   
${\rm E}\{|\epsilon/c_H^{1/2}|\Lambda(\epsilon/c_H^{1/2})\}< \infty.$
\item[(iii)] There exists function $\Lambda^*$ satisfying
\begin{equation*} 
|\Lambda(e+es)-\Lambda(e)| \le \Lambda^*(e)s, \, e \in \R^1, \, s>0,
\end{equation*}
where ${\rm E}\log^+\{\Lambda^*(\epsilon/c_H^{1/2})\}< \infty.$
\end{itemize}
\bigskip 

\color{black}

Defining the {\it score function factor}  
$$
\sigma^2(H)\coloneqq 4\,\,\mbox{Var}\{H(\epsilon/c_H^{1/2})\}/[{\mathrm E}\{(\epsilon/c_H^{1/2})
\d{H}(\epsilon/c_H^{1/2})\}]^2,
$$
and the matrix  
$$
\G\coloneqq {\mathrm E}\{\d{\mbf v}_1(\bth_{0H})\d{\mbf v}_1^{\prime}(\bth_{0H})/v_1^2(\bth_{0H})\}.
$$
Then the following result on the asymptotic distribution holds (Mukherjee (2008)). 
\begin{theorem}\label{thm.Mest}
Suppose that Assumptions (A) and (B1)-(B2) hold.  Then~$n^{1/2}(\hat{\bth}_n-\bth_{0H})$ is asymptotically normal with mean $\0$ and covariance $\sigma^2(H)\G^{-1}$ as~$n\to~\!\infty$. 
\end{theorem}

{\bf Remark 1.} Note that the values of the coefficients $c_H$ in Assumption~(B1) are 
$c_H = {\mathrm E}(\epsilon^2)$ for the QMLE and~$c_H = \left({\mathrm E}|\epsilon|\right)^2$ for the LAD. For the Huber, $\mu$-, Cauchy, and other scores, $c_H$ does not have a closed-form expression but the corresponding numerical values can be computed from~(\ref{H1}) for various error distributions as follows. Fix a large positive integer $I$ and generate $\{\epsilon_i; 1 \le i \le I\}$ from the error distribution. Then, using the bisection method on $c >0 $,  solve the equation 
$$
(1/I)\sum_{i=1}^I\left\lbrace H\left(\epsilon_i/c^{1/2}\right)\right\rbrace -1=0.
$$
In Table \ref{Tab.cH}, we provide $c_H$ for some further  error distributions and score functions such as Huber's $k$-score with $k=1.5$ and the $\mu$-estimator  with $\mu=3$, which are used in simulations and data analysis in subsequent  sections. 
\begin{table}[h!]
\caption{Values of $c_H$ for various M-estimators (Huber, $\mu$-, Cauchy) under normal, double-exponential (DE), logistic, $t(3)$, and $t(2.2)$ error distributions.\vspace{2mm}}\label{Tab.cH}
\centering
\begin{tabular}{c c c c}
\hline
\hline
   &    &  & \vspace{-3mm}  \\
   & Huber  & $\mu$-estimator & Cauchy\vspace{1mm} \\
\hline
Normal & 0.825 & 1.692 & 0.377\\
DE &  0.677 & 1.045 & 0.207 \\
Logistic &  0.781 & 1.487 & 0.316 \\
$t(3)$ & 0.533 & 0.850 & 0.172 \\
$t(2.2)$ & 0.204 & 0.274 & 0.053 \\
\hline
\hline
\end{tabular}
\end{table}

\section{Bootstrapping M-estimators}\label{sec.BootMest}

Let $\{w_{nt}; 1 \le t \le n, n \ge 1\}$
be a triangular array of random variables such that (i) for each~$n \ge 1$, 
$\{w_{nt}; 1 \le t \le n\}$ are exchangeable and independent of 
$\{X_t; t \ge 1\}$ and $\{\epsilon_t; t \ge 1\}$, and (ii)  $w_{nt} \ge 0$ and ${\mathrm E}(w_{nt})=1$ for all $t \ge 1$.  
 Based on these weights~$w_{nt}$, a bootstrap estimate $\hat{\bth}_{*n}$ is defined as a solution of~$\widehat{\mbf M}^*_{n, H} (\bth) = \0$, where
\begin{equation}\label{defnb}
\widehat{\mbf M}^*_{n, H} (\bth) \coloneqq  \s w_{nt}\bigg\{1-H\{X_t/\h{v}_t^{1/2}(\bth)\}\bigg\}\{\d{\h{\mbf v}}_t(\bth)/\h{v}_t(\bth)\}.
\end{equation}
{\bf Examples.} From various  available choices of the bootstrap weights, we consider, for the sake of comparison,  the following three bootstrapping schemes. 
\begin{enumerate}
\item[(i)] {\bf Scheme M.} The sequence of weights $\{w_{n1}, \ldots, w_{nn}\}$ has a multinomial $(n, 1/n, \ldots, 1/n)$ distribution, which is essentially the
classical paired bootstrap. 
\item[(ii)] {\bf Scheme E.} The weights are of the form  $w_{nt}=(n E_t)/\sum_{i=1}^n E_i$, where $\{E_t\}$ are i.i.d.\ exponential  with mean $1$. 
\item[(iii)] {\bf Scheme U.} The weights are of the form  $w_{nt}=(n U_t)/\sum_{i=1}^n U_i$,
where $\{U_t\}$ are i.i.d.\ uniform  on $(0.5, 1.5)$. 
\end{enumerate}  

A host of other bootstrap methods in the literature
are special cases of the above 
 formulation. Such general formulation of weighted bootstrap offers a
unified way of studying several bootstrap schemes simultaneously. See, for instance, Chatterjee and Bose (2005) for details in different contexts.

We assume that the weights satisfy the following basic conditions (Conditions BW of Chatterjee and Bose (2005)) where
$\sigma_n^2= {\rm Var}(w_{ni})$ and $k_3>0$ is a constant:
\begin{equation}\label{B2}
{\mathrm E}(w_{n1})=1, \,\, 0<k_3<\sigma_n^2=o(n), 
\,\,\mbox{and}\,\,\mbox{Corr}\,(w_{n1}, w_{n2})=O(1/n).
\end{equation}
We also assume additional smoothness and moment conditions:

{\bf Assumptions (B')} $H(x)$ is twice differentiable at all but a finite number of points and for some $\delta>2$,
${\mathrm E}[H(\epsilon/c_H^{1/2})]^{\delta} < \infty$.\bigskip

Under (\ref{B2}) and Assumptions (A), (B) and (B'), the weighted bootstrap is asymptotically valid (Mukherjee (2020)).\medskip 

\begin{theorem}\label{thm.BootMest} 
Suppose that Assumptions (A), (B), (B') and (\ref{B2}) hold. Then for almost all data as~$n \rightarrow \infty$, 
\begin{equation}\label{an} 
\sigma_n^{-1} n^{1/2}(\hat{\bth}_{*n}- \hat{\bth}_n)\rightarrow {\cal N}(\0, \sigma^2(H)\G^{-1}).
\end{equation}
\end{theorem}
Since $0<1/\sigma_n<1/\sqrt{k_3}$, the rate of convergence of the bootstrap estimator is the same as that of the original M-estimator. The standard deviation of the weights $\{\sigma_n\}$ in the denominator of the scaling reflects the impact of the chosen weights. 

The distributional result of (\ref{an}) is useful for constructing  confidence intervals  for the GARCH parameters. Let $B$ be the number of bootstrap replicates. Consider the true value~$\gamma_0$ of a generic parameter (either $\omega_0$, $\alpha_{0i}$, or $\beta_{0j}$) and let 
$\hat{\gamma}_n$ and $\hat{\gamma}_{*nb}$ denote its \linebreak M-estimator and $b$-th bootstrap estimator ($1 \le b \le B$), respectively. Let $\gamma_{0H}$ denote the corresponding  transformed parameter  (either $c_H\omega_0$, $c_H\alpha_{0i}$, or $\beta_{0j}$; see \eqref{thetanoth}); the value of this  $\gamma_{0H}$ is known in simulation experiments.

Using the approximation of 
$\sqrt{n} (\h{\gamma}_{n} - \gamma_{0H})$ by $\sigma_n^{-1} n^{1/2}(\h{\gamma}_{*n} - \h{\gamma}_{n})$, the bootstrap confidence interval (with confidence level $(1-\alpha)$) for $\gamma_{0H}$ is of the form  
\begin{equation}\label{bci}
\Big[\h{\gamma}_{n}-n^{-1/2} \{\sigma_n^{-1} n^{1/2}(\h{\gamma}_{*n, \alpha/2} - \h{\gamma}_{n})\}, \ \h{\gamma}_{n}+n^{-1/2} \{\sigma_n^{-1} n^{1/2}(\h{\gamma}_{*n, 1 - \alpha/2} - \h{\gamma}_{n})\}\Big]
\end{equation}  
where $\h{\gamma}_{*n, \alpha/2}$ is the $\alpha/2$-th quantile of the numbers $\{\hat{\gamma}_{*nb}, 1 \le b \le B\}$. Consequently, the bootstrap coverage probability is evaluated by the proportion of intervals of the form \eqref{bci} containing 
$\gamma_{0H}$.   

The asymptotic normality result of  Theorem ~\ref{thm.Mest} also yields a confidence interval (with confidence level $(1-\alpha)$) for $\gamma_{0H}$; we call it the normal confidence interval. 
This is of the form 
\begin{equation}\label{nci}
\Big[\h{\gamma}_{n}-n^{-1/2} \hat{d} z_{1-\alpha/2}, \ \h{\gamma}_{n}+n^{-1/2} \hat{d} z_{1-\alpha/2}\Big]
\end{equation}   
where $\hat{d}^2$ is the estimated variance of $\h{\gamma}_{n}$ obtained as the appropriate diagonal entry of the estimator of 
$\sigma^2(H)\G^{-1}$ and $ z_{1-\alpha/2}$ is the ($1-\alpha/2$)-th quantile of the standard normal distribution. 

In Section \ref{sec.simBootMest}, we compare the accuracy of the bootstrap-based and normal confidence intervals \eqref{bci} and \eqref{nci}. 

\section{Computational issues}\label{sec.comMest}   
\setcounter{equation}{0}
This section is devoted to  the detail implementation of an iteratively re-weighted algorithm for the computation of M-estimates proposed in Mukherjee (2020). In particular, we highlight the~$\mu$-  and Cauchy-estimates, since  their asymptotic distributions are derived under mild moment assumptions. We also consider the bootstrap estimators based on the corresponding score functions. 

\subsection{Computation of the M-estimates}
		
For the convenience of writing, let $\alpha(c)\coloneqq{\mathrm E}[H(c \epsilon)]$ for $c>0$. Using a Taylor expansion of~$\widehat{\mbf M}_{n, H}$, we obtain the following recursion yielding the updated estimate 
$\tilde{\bth}$ of $\hat{\bth}_n$ as a function of the current one $\check{\bth}$, say, 
\begin{equation}\label{just3}
\tilde{\bth}= {\check{\bth}}+\{\dot{\alpha}(1)/2\}^{-1}\Big[\sum_{t=1}^n\d{\h{\mbf v}}_t({\check{\bth}})\d{\h{\mbf v}}_t^{\prime}({\check{\bth}})/\hat{v}_t^2({\check{\bth}}) \Big]^{-1} 
\sum_{t=1}^n \Big\{H\{X_{t}/\hat{v}_{t}^{1/2}({\check{\bth}})\} -1\Big\}\{\d{\h{\mbf v}}_{t}({\check{\bth}})/\hat{v}_{t}({\check{\bth}}\}, 
\end{equation}
where $\dot{\alpha}(1)={\mathrm E}\{\epsilon \dot{H}(\epsilon)\}$ (this expectation exists under the smoothness conditions in Assumption (B)). We now discuss two aspects regarding implementation of the algorithm in~\eqref{just3}. First, the initial value of $\check{\bth}$ for the iteration, in principal, should be a $\sqrt{n}$-consistent estimator of $\bth_{0H}$. However, we observe in our extensive simulation study that irrespective of the choice of the QMLE, LAD, $\bth_0$ or even values very different from $\bth_{0H}$ as initial estimates, only few iterations are needed for the convergence to the same estimates. Second, we cannot, in general, estimate $\dot{\alpha}(1)$ from the data  using the GARCH residuals $\{X_t/ \hat{v}_t^{1/2}(\hat{\bth}_n)\}$ as 
they are close to $\{\epsilon_t/c_H^{1/2}\}$, an unknown multiplicative factor of the errors.   
Therefore, we use ad-hoc techniques such as simulating 
$\{\tilde{\epsilon}_t; 1 \le t \le n\}$ from ${\mathcal N} (0, 1)$ or standardized double exponential distributions and then use  
$n^{-1} \s \tilde{\epsilon} \dot{H}(\tilde{\epsilon})$
to carry out the iterations. Note that if the iteration in (\ref{just3}) converges, then
$\tilde{\bth}-\check{\bth}\approx \0$, hence $\widehat{\mbf M}_{n, H} (\check{\bth}) \approx \0$, and~$\tilde{\bth}$ is the desired~$\hat{\bth}_n$. Based on our extensive simulation study and real data analysis, this algorithm appears to be  robust enough to converge to the same value of $\hat{\bth}_n$ irrespective of  the evaluations  of the unknown value of  
$\dot{\alpha}(1)$ used in the computation. 

In the following examples, we discuss~(\ref{just3}) when specialized to the M-estimates computed in this paper.  
\bigskip

\noindent (a) {\bf QMLE}. Here $H(x) = x^2$ and $\alpha(c)=c^2{\mathrm E}(\epsilon^2)$. Hence $\dot{\alpha}(1)/2={\mathrm E}(\epsilon^2)$ and  \eqref{just3} takes the form
\begin{eqnarray*}
\tilde{\bth}&=&{\check{\bth}}
+\Big \{{\mathrm E}(\epsilon^2)\Big\}^{-1} \Big[\sum_{t=1}^n\Big\{\d{\h{\mbf v}}_t({\check{\bth}})\d{\h{\mbf v}}_t^{\prime}({\check{\bth}})/\hat{v}_t^2({\check{\bth}})\Big\} \Big]^{-1}
\sum_{t=1}^n \Big[\{X_{t}^2/\hat{v}_t({\check{\bth}})\}-1\Big]\{\d{\h{\mbf v}}_t({\check{\bth}})/\hat{v}_t({\check{\bth}})\}.
\end{eqnarray*}
With
 $
W_{t} = 1/\hat{v}_t^2({\tilde{\bth}_{(r)}})$, $x_{t} = \d{\h{\mbf v}}_t({\tilde{\bth}_{(r)}})$, and $ y_{t}=X_t^2- \hat{v}_t({\tilde{\bth}_{(r)}})$, 
$\tilde{\bth}_{(r+1)}$  (iteration $r+1$) thus 
is to be computed  as 
\begin{equation*}
\tilde{\bth}_{(r+1)} =\tilde{\bth}_{(r)} + \Big \{{\mathrm E}(\epsilon^2)\Big\}^{-1} \left\{\sum_{t}W_{t}x_{t}x_{t}'\right\}^{-1}\left\{\sum_{t}W_{t}x_{t}y_{t}\right\}.
\end{equation*} 
Note that when ${\mathrm E}(\epsilon^2)=1$, this coincides with the formula obtained through the BHHH algorithm proposed by Berndt et al.~(1974).\medskip 

\noindent (b) {\bf LAD.} Here $H(x) = |x|$ and $\alpha(c)=c {\mathrm E}|\epsilon|$. Hence $\dot{\alpha}(1)={\mathrm E}|\epsilon|$ and  \eqref{just3} takes the form
\begin{eqnarray*}
\tilde{\bth}&=&{\check{\bth}}
+\{2/{\mathrm E}|\epsilon|\}\Big[\sum_{t=1}^n\Big\{\d{\h{\mbf v}}_t({\check{\bth}}))\d{\h{\mbf v}}_t^{\prime}({\check{\bth}}))/\hat{v}_t^2({\check{\bth}}))\Big\} \Big]^{-1}
\sum_{t=1}^n \Big[|X_{t}|/\hat{v}_t^{1/2}({\check{\bth}}))-1\Big]\{\d{\h{\mbf v}}_t({\check{\bth}}))/\hat{v}_t({\check{\bth}}))\}\\
&=&{\check{\bth}}+
\{2/{\mathrm E}|\epsilon|\}\Big[\sum_{t=1}^n\Big\{\d{\h{\mbf v}}_t({\check{\bth}})\d{\h{\mbf v}}_t^{\prime}({\check{\bth}})/\hat{v}_t^2({\check{\bth}})\Big\} \Big]^{-1}
\sum_{t=1}^n \Big\{|X_{t}|-\hat{v}_t^{1/2}({\check{\bth}})\Big\}\{\d{\h{\mbf v}}_t({\check{\bth}})/\hat{v}^{3/2}_t({\check{\bth}})\}\\
&=& {\check{\bth}}+\{2/{\mathrm E}|\epsilon|\}
\Big[\sum_{t=1}^n\Big\{\d{\h{\mbf v}}_t({\check{\bth}})\d{\h{\mbf v}}_t^{\prime}({\check{\bth}})/\hat{v}_t^2({\check{\bth}})\Big\} \Big]^{-1}
\sum_{t=1}^n \Big\{ \hat{v}_t^{1/2}({\check{\bth}})(|X_{t}|-\hat{v}_t^{1/2}({\check{\bth}}))\Big\} \{\d{\h{\mbf v}}_t({\check{\bth}})/\hat{v}^{2}_t({\check{\bth}})\}.
\end{eqnarray*}
With
$
W_{t} = 1/\hat{v}_t^2({\tilde{\bth}_{(r)}})$, $x_{t} = \d{\h{\mbf v}}_t({\tilde{\bth}_{(r)}})$, and $y_{t}=\hat{v}_t^{1/2}({\tilde{\bth}_{(r)}})(|X_{t}|-\hat{v}_t^{1/2}({\tilde{\bth}_{(r)}}))$, 
${\tilde{\bth}_{(r+1)}}$  (iteration~$r+~\!1$) thus is to be computed  as
\begin{equation*}
\tilde{\bth}_{(r+1)} =\tilde{\bth}_{(r)} +\{2/{\mathrm E}|\epsilon|\}\left\{\sum_{t}W_{t}x_{t}x_{t}'\right\}^{-1}\left\{\sum_{t}W_{t}x_{t}y_{t}\right\}.
\end{equation*} 

\noindent (c) {\bf Huber.}
Here $H(x) = x^2I(|x|\leq k)+k|x|I(|x|>k)$ and   
$$\alpha(c)={\mathrm E}\left[(c\epsilon)^2I(|c\epsilon|\leq k)+k|c\epsilon|I(|c\epsilon|>k)\right].$$
Hence 
\begin{equation*}
\dot{\alpha}(1)={\mathrm E}\left[2\epsilon^2I(|\epsilon|\leq k)+k|\epsilon|I(|\epsilon|>k)\right]
\end{equation*} and  \eqref{just3} takes the form  
\begin{eqnarray}
\tilde{\bth}&=&{\check{\bth}}
-\Big \{\dot{\alpha}(1)/2\Big\}^{-1} \Big[\sum_{t=1}^n\Big\{\frac{\d{\h{\mbf v}}_t({\check{\bth}})\d{\h{\mbf v}}_t^{\prime}({\check{\bth}})}{\hat{v}_t^2({\check{\bth}})}\Big\} \Big]^{-1} \nn \\
&& \times \sum_{t=1}^n \left[1-\frac{X_{t}^2}{\hat{v}_t({\check{\bth}})}I\left(\frac{|X_t|}{\hat{v}_t^{1/2}({\check{\bth}})}\leq k\right)-k\frac{|X_t|}{\hat{v}_t^{1/2}({\check{\bth}})}I\left(\frac{|X_t|}{\hat{v}_t^{1/2}({\check{\bth}})}> k\right)\right]\left\lbrace\frac{\d{\h{\mbf v}}_t({\check{\bth}})}{\hat{v}_t({\check{\bth}})}\right\rbrace. \nonumber
\end{eqnarray}
With
 $
W_{t} = 1/\hat{v}_t^2({\tilde{\bth}_{(r)}})$, $x_{t} = \d{\h{\mbf v}}_t({\tilde{\bth}_{(r)}})
$ 
and
$$
y_{t}=X_t^2I\left(|X_t|/\hat{v}_t^{1/2}({\tilde{\bth}_{(r)}})\leq k\right)+k|X_t|\hat{v}_t^{1/2}({\tilde{\bth}_{(r)}})I\left(|X_t|/\hat{v}_t^{1/2}({\tilde{\bth}_{(r)}})> k\right)-\hat{v}_t({\tilde{\bth}_{(r)}}),
$$
${\tilde{\bth}_{(r+1)}}$  (iteration~$r+~\!1$) thus is to be computed  as
\begin{equation*}
\tilde{\bth}_{(r+1)} =\tilde{\bth}_{(r)} + \Big \{\dot{\alpha}(1)/2\Big\}^{-1} \left\{\sum_{t}W_{t}x_{t}x_{t}'\right\}^{-1}\left\{\sum_{t}W_{t}x_{t}y_{t}\right\}.
\end{equation*} 

\noindent (d) {\bf $\boldsymbol\mu$-estimator.}
Here $H(x) = \mu |x|/(1+|x|)$ and $\alpha(c)=\mu-\mu {\mathrm E}\left[1/(1+|c\epsilon|)\right]$. Hence 
$$
\dot{\alpha}(1)=\mu {\mathrm E}\left[|\epsilon|/(1+|\epsilon|)^2\right]
$$ 
and  \eqref{just3} takes the form  
\begin{equation*}
\tilde{\bth}={\check{\bth}}+\left\lbrace\frac{\mu}{2}{\mathrm E}\left[\frac{|\epsilon|}{(1+|\epsilon|)^2}\right]\right\rbrace^{-1} \Big[\sum_{t=1}^n\Big\{\frac{\d{\h{\mbf v}}_t({\check{\bth}})\d{\h{\mbf v}}_t^{\prime}({\check{\bth}})}{\hat{v}_t^2({\check{\bth}})}\Big\} \Big]^{-1} 
\sum_{t=1}^n \left[\frac{\mu|X_{t}|}{\hat{v}_t^{1/2}({\check{\bth}})+|X_t|}-1\right]\left\lbrace\frac{\d{\h{\mbf v}}_t({\check{\bth}})}{\hat{v}_t({\check{\bth}})}\right\rbrace.
\end{equation*}
With
$
W_{t} = 1/\hat{v}_t^2({\tilde{\bth}_{(r)}})$, $x_{t} = \d{\h{\mbf v}}_t({\tilde{\bth}_{(r)}})$, and $y_{t}=\dfrac{\mu|X_t|\hat{v}_t({\tilde{\bth}_{(r)}})}{\hat{v}_t^{1/2}({\tilde{\bth}_{(r)}})+|X_t|}-\hat{v}_t({\tilde{\bth}_{(r)}}),
$ 
${\tilde{\bth}_{(r+1)}}$  (iteration~$r+~\!1$) thus is to be computed  as
\begin{equation*}
\tilde{\bth}_{(r+1)} =\tilde{\bth}_{(r)} + \left\lbrace\frac{\mu}{2}{\mathrm E}\left[\frac{|\epsilon|}{(1+|\epsilon|)^2}\right]\right\rbrace^{-1} \left\{\sum_{t}W_{t}x_{t}x_{t}'\right\}^{-1}\left\{\sum_{t}W_{t}x_{t}y_{t}\right\}.
\end{equation*} 

\noindent (e) {\bf Cauchy-estimator.}
Here $H(x) = 2x^2/(1+x^2)$ and $\alpha(c)={\mathrm E}\left[2c^2\epsilon^2/(1+c^2\epsilon^2)\right]$. Hence 
$$
\dot{\alpha}(1)={\mathrm E}\left[4\epsilon^2/(1+\epsilon^2)^2\right]
$$ and  
\begin{equation*}
\tilde{\bth}={\check{\bth}}
-\left\lbrace2{\mathrm E}\left[\frac{\epsilon^2}{(1+\epsilon^2)^2}\right]\right\rbrace^{-1} \Big[\sum_{t=1}^n\Big\{\frac{\d{\h{\mbf v}}_t({\check{\bth}})\d{\h{\mbf v}}_t^{\prime}({\check{\bth}})}{\hat{v}_t^2({\check{\bth}})}\Big\} \Big]^{-1} 
\sum_{t=1}^n \left[1-\frac{2X_t^2}{\hat{v}_t({\check{\bth}})+X_t^2}\right]\left\lbrace\frac{\d{\h{\mbf v}}_t({\check{\bth}})}{\hat{v}_t({\check{\bth}})}\right\rbrace.
\end{equation*}
With
 $W_{t} = 1/\hat{v}_t^2({{\tilde{\bth}_{(r)}}})$, $x_{t} = \d{\h{\mbf v}}_t({{\tilde{\bth}_{(r)}}})$, and $y_{t}=\dfrac{2X_t^2\hat{v}_t({\tilde{\bth}_{(r)}})}{\hat{v}_t({\tilde{\bth}}_{(r)})+X_t^2} - \hat{v}_t(\tilde{\bth}_{(r)})$,
${\tilde{\bth}_{(r+1)}}$  (iteration~$r+~\!1$) thus is to be computed  as
\begin{equation*}
\tilde{\bth}_{(r+1)} =\tilde{\bth}_{(r)} + \left\lbrace2{\mathrm E}\left[\frac{\epsilon^2}{(1+\epsilon^2)^2}\right]\right\rbrace^{-1} \left\{\sum_{t}W_{t}x_{t}x_{t}'\right\}^{-1}\left\{\sum_{t}W_{t}x_{t}y_{t}\right\}.
\end{equation*} 

\subsection{Computation of  the bootstrap M-estimates}
The relevant function here is $\widehat{\mbf M}^*_{n, H}(\bth)$ defined in (\ref{defnb})
and the bootstrap estimate $\hat{\bth}_{*n}$ can be computed from the current one ${\check{\bth}_*}$, say, using the updating equation
\begin{eqnarray}\label{just4}
\tilde{\bth}_* &= &{\check{\bth}_*}-\{2/\dot{\alpha}(1)\}\Big[\sum_{t=1}^n\ w_{nt}\Big\{\d{\h{\mbf v}}_t({\check{\bth}_*})\d{\h{\mbf v}}_t^{\prime}({\check{\bth}_*})/\hat{v}_t^2({\check{\bth}_*})\Big\} \Big]^{-1} \nonumber\\
& & \qquad \times \sum_{t=1}^n w_{nt}\Big\{1-H\{X_{t}/\hat{v}_{t}^{1/2}({\check{\bth}_*})\} \Big\}\{\d{\h{\mbf v}}_{t}({\check{\bth}_*})/\hat{v}_{t}({\check{\bth}_*})\},
\end{eqnarray}  
\color{black}where the M-estimate $\hat{\bth}_{n}$ obtained via iteration process \eqref{just3} is chosen as the initial value. 

We remark that the weighted bootstrap is more computational friendly and easy-to-implement than the commonly-applied residual bootstrap (see, e.g., Jeong (2017)) for GARCH models, since it avoids computation of residuals at each iteration. In particular, one simply needs to generate weights once to compute a bootstrap estimate.

\color{black}


\section{Monte Carlo comparison of 
performance}\label{sec.simMest} 

To compare the finite-sample performance of various M-estimators via their bias and men squared errors (MSE), we simulate $n$ observations from GARCH models with specific choices of parameters and error distributions and compute the resulting M-estimates based on various score functions. This procedure is replicated $R$-times to enable the estimation of bias and MSE. For instance, with $p=1=q$, 
let $\hat{\bth}_n=(\hat{\omega}_r, \hat{\alpha}_r, \hat{\beta}_r)^{\prime}$ be the
M-estimator of~$\bth_0=(\omega_0, \alpha_{01}, \beta_{01})^{\prime}$ based on the score function $H$ at the $r$-th replication, $1 \le r \le R$. However, $(\hat{\omega}_r, \hat{\alpha}_r, \hat{\beta}_r)$  is a consistent estimator of~$(c_H\omega_0, c_H\alpha_0, \beta_0)$, where $c_H$ depends   on  the score function and the underlying error distribution (which are known in a simulation scenario).  Therefore, we compare the performance at a specified error distribution across various score functions in terms of the {\it adjusted bias} and {\it adjesusted MSEs} defined by
$$
E(\hat{\omega}/c_H-\omega_0), \,\,\, E(\hat{\alpha}/c_H-\alpha_0), \,\,\, E(\hat{\beta}-\beta_0)
$$
and 
$$
E(\hat{\omega}/c_H-\omega_0)^2, \,\,\, E(\hat{\alpha}/c_H-\alpha_0)^2,  \,\,\, E(\hat{\beta}-\beta_0)^2.
$$
We consider $R$ replicates of
$$
(\hat{\omega}_r/c_H-\omega_0, \hat{\alpha}_r/c_H-\alpha_0, \hat{\beta}_r-\beta_0)^\prime
$$        
and use the following quantities to estimate the {\it adjusted biases} 
\begin{equation}\label{sbias}
R^{-1}\sum_{r=1}^R \{\hat{\omega}_r/c_H-\omega_0\}, \ \ R^{-1}\sum_{r=1}^R \{\hat{\alpha}_r/c_H-\alpha_0\}, \ \ R^{-1}\sum_{r=1}^R \{\hat{\beta}_r-\beta_0\}
\end{equation}
and the {\it adjusted MSEs} 
$$
R^{-1}\sum_{r=1}^R \{\hat{\omega}_r/c_H-\omega_0\}^2, \ \ R^{-1}\sum_{r=1}^R \{\hat{\alpha}_r/c_H-\alpha_0\}^2, \ \ R^{-1}\sum_{r=1}^R \{\hat{\beta}_r-\beta_0\}^2.
$$

We consider the GARCH~(1, 1) model in Section~\ref{GARCH11} and higher-order GARCH (2,1) and GARCH (1,2) models  in Sections~\ref{GARCH21} and~\ref{GARCH12} respectively. Section~\ref{missp} considers a case of misspecified GARCH orders. 

\subsection{GARCH~(1, 1) models}\label{GARCH11}

In Tables \ref{Huber.k} and \ref{mu.mu}, we report the adjusted biases and MSEs of  the Huber and $\mu$-type 
M-estimators to guide our choice of the  tuning parameters $k$ and $\mu$. The underlying data-generating process (DGP) is the GARCH~(1, 1) model with $\bth_0=(1.65\times 10^{-5}, 0.0701, 0.901)^{\prime}$, 
under three types of innovation distributions:   normal, double exponential, and logistic. 
 The  sample size is $n=1000$ and we used $R=150$ replications. 

\begin{table}[!htbp]
\caption{The adjusted bias and MSE of  Huber estimators for various values of $k$  under  a GARCH~(1,1) model with various error distributions (Normal, double exponential, logistic); sample size~$n =~\!1000$; $R = 150$ replications.}\label{Huber.k} \vspace{3mm}
\centering
\begin{tabular}{c c c c c c c c} \hline \hline
 &\multicolumn{3}{c}{\textbf{ }} &&\multicolumn{3}{c}{\textbf{ }\vspace{-1.5mm}} \\ 
  &\multicolumn{3}{c}{\textbf{adjusted bias}} &&\multicolumn{3}{c}{\textbf{adjusted MSE}\vspace{1mm}} \\ \cline{2-4} \cline{6-8}   & $\omega$ &$\alpha$&$\beta$&  &$\omega$ &$\alpha$&$\beta$ \\ \hline
 Normal  &&& & &&& \\
  k=1 & 1.03$\times 10^{-5}$ & -2.44$\times 10^{-3}$ & -1.96$\times 10^{-2}$ & & 2.62$\times 10^{-10}$ & 4.20$\times 10^{-4}$ & 1.54$\times 10^{-3}$ \\
 k=1.5 & 1.22$\times 10^{-5}$ & 2.47$\times 10^{-3}$ & -1.98$\times 10^{-2}$& & 3.33$\times 10^{-10}$ & 4.55$\times 10^{-4}$ & 1.58$\times 10^{-3}$  \\
 k=2.5 & 1.14$\times 10^{-5}$ & -4.33$\times 10^{-4}$ & -2.02$\times 10^{-2}$&  & 3.10$\times 10^{-10}$ & 3.71$\times 10^{-4}$ & 1.58$\times 10^{-3}$ \\ \hline
 DE  &&& & &&& \\
 k=1 & 7.24$\times 10^{-6}$ & 1.29$\times 10^{-3}$ & -1.57$\times 10^{-2}$&  & 1.87$\times 10^{-10}$ & 4.65$\times 10^{-4}$ & 1.58$\times 10^{-3}$ \\
 k=1.5 & 7.32$\times 10^{-6}$ & 1.67$\times 10^{-3}$& -1.63$\times 10^{-2}$& & 2.00$\times 10^{-10}$ & 4.82$\times 10^{-4}$ & 1.68$\times 10^{-3}$ \\
k=2.5 & 8.27$\times 10^{-6}$ & 2.94$\times 10^{-3}$ & -1.92$\times 10^{-2}$ & & 2.79$\times 10^{-10}$ & 5.60$\times 10^{-4}$ & 2.22$\times 10^{-3}$  \\ \hline
Logistic  &&& & &&& \\
 k=1 & 9.87$\times 10^{-6}$ & 2.15$\times 10^{-3}$ & -2.03$\times 10^{-2}$& & 3.18$\times 10^{-10}$ & 5.25$\times 10^{-4}$ & 2.28$\times 10^{-3}$  \\
k=1.5 & 1.00$\times 10^{-5}$ & 2.04$\times 10^{-3}$ &-2.04$\times 10^{-2}$& & 3.11$\times 10^{-10}$ & 4.89$\times 10^{-4}$ & 2.22$\times 10^{-3}$ \\
 k=2.5 & 1.06$\times 10^{-5}$& 2.18$\times 10^{-3}$ & -2.16$\times 10^{-2}$& &3.18$\times 10^{-10}$& 4.84$\times 10^{-4}$ & 2.17$\times 10^{-3}$ \\ \hline\hline
\end{tabular}
\end{table}

\begin{table}[!htbp]
\caption{The adjusted bias and MSE of $\mu$-estimators for  various values of $\mu$   under  a GARCH~(1,1) model with various error distributions (Normal, double exponential, logistic); sample size~$n =~\!1000$; $R = 150$ replications.}\label{mu.mu} \vspace{3mm}
\begin{tabular}{c c c c c c c c} \hline \hline
 &\multicolumn{3}{c}{\textbf{ }} &&\multicolumn{3}{c}{\textbf{ }\vspace{-1.5mm}} \\ 
 &\multicolumn{3}{c}{\textbf{adjusted bias}} &&\multicolumn{3}{c}{\textbf{adjusted MSE}\vspace{1mm}} \\ \cline{2-4} \cline{6-8}   & $\omega$ &$\alpha$&$\beta$&  &$\omega$ &$\alpha$&$\beta$ \\ \hline
Normal  &&& & &&& \\
 $\mu$=2 & 1.17$\times 10^{-5}$ & 2.97$\times 10^{-3}$ &-2.13$\times 10^{-2}$ & & 4.05$\times 10^{-10}$ & 6.73$\times 10^{-4}$ & 2.16$\times 10^{-3}$  \\
$\mu$=2.5 & 1.14$\times 10^{-5}$ & 1.80$\times 10^{-3}$ & -2.12$\times 10^{-2}$ & & 3.77$\times 10^{-10}$ & 5.71$\times 10^{-4}$ & 2.04$\times 10^{-3}$ \\
$\mu$=3 & 1.14$\times 10^{-5}$ & 1.36$\times 10^{-3}$ & -2.11$\times 10^{-2}$ & & 3.68$\times 10^{-10}$ & 5.21$\times 10^{-4}$ & 1.97$\times 10^{-3}$   \\ \hline
DE  &&& & &&& \\
 $\mu$=2 & 7.39$\times 10^{-6}$ & 2.23$\times 10^{-3}$ & -1.49$\times 10^{-2}$& &2.74$\times 10^{-10}$ & 7.20$\times 10^{-4}$ & 2.21$\times 10^{-3}$ \\
 $\mu$=2.5 & 7.36$\times 10^{-6}$ & 1.50$\times 10^{-3}$ & -1.52$\times 10^{-2}$&& 2.68$\times 10^{-10}$ & 6.56$\times 10^{-4}$ & 2.16$\times 10^{-3}$   \\
 $\mu$=3 & 7.40$\times 10^{-6}$ & 1.25$\times 10^{-3}$ & -1.53$\times 10^{-2}$& & 2.62$\times 10^{-10}$ & 6.17$\times 10^{-4}$ & 2.09$\times 10^{-3}$ \\ \hline
Logistic  &&& & &&& \\
 $\mu$=2 & 7.73$\times 10^{-6}$ & 2.22$\times 10^{-3}$ & -1.37$\times 10^{-2}$& & 2.45$\times 10^{-10}$ & 6.79$\times 10^{-4}$ & 1.99$\times 10^{-3}$  \\
$\mu$=2.5 & 7.66$\times 10^{-6}$ & 9.77$\times 10^{-4}$ & -1.41$\times 10^{-2}$& & 2.48$\times 10^{-10}$ & 5.88$\times 10^{-4}$ & 1.97$\times 10^{-3}$ \\
 $\mu$=3 & 7.72$\times 10^{-6}$ & 5.99$\times 10^{-4}$ & -1.42$\times 10^{-2}$& & 2.54$\times 10^{-10}$ & 5.44$\times 10^{-4}$ & 1.94$\times 10^{-3}$  \\ \hline\hline
\end{tabular}
\end{table} 

Results from Table \ref{Huber.k} and Table~\ref{mu.mu} reveal that the adjusted bias and MSE of Huber's $k$-estimator and the $\mu$-estimator do not vary much with $k$ and $\mu$. Therefore,~$k=1.5$ and~$\mu=3$ are chosen for subsequent computations. Notice also that the minimum bias and MSE are obtained for the~$\mu$-estimator with $\mu=3$ in most cases.


\subsection{GARCH~(2, 1) models}\label{GARCH21}

In this section,we consider GARCH~(2, 1) models with five types of innovation distributions: the normal, double exponential, logistic, and Student' $t$ with $3$  and $2.2$ degrees of freedom (denoted by $t(3)$ and $t(2.2)$). The sample size is still $n=1000$ and $R= 1000$ replications were generated from the GARCH~(2, 1) model with parameter
$$
\boldsymbol\theta_0 = (4.46\times 10^{-6}, 0.0525, 0.108, 0.832)^\prime,
$$ 
and this choice is motivated by the QMLE computed from the FTSE~100 dataset analyzed in Section~\ref{FTSE100}
using the R package {\tt fGarch}.   


\begin{table}[t]
\caption{The adjusted bias and MSE of   M-estimators for GARCH~(2, 1) models under various error distributions (Normal, double exponential, logistic, $t(3)$, $t(2.2)$); sample size~$n =~\!1000$; $R = 1000$ replications.\vspace{3mm}}\label{Sim.GARCH21}
\centering
\scriptsize  
\begin{tabular}{c c c c c c c c c c}\hline\hline 
&\multicolumn{4}{c}{} &&\multicolumn{4}{c}{ \vspace{-2mm}}\\
 &\multicolumn{4}{c}{\textbf{adjusted bias}} &&\multicolumn{4}{c}{\textbf{adjusted MSE}\vspace{1.5mm}} \\
  \cline{2-5} \cline{7-10} & $\omega$ &$\alpha_1$& $\alpha_2$& $\beta$&  &$\omega$ &$\alpha_1$ & $\alpha_2$&$\beta$ \\ \hline
\textbf{Normal} &&&&&&&&&  \\
QMLE           & 3.55$\times 10^{-6}$ & 1.88$\times 10^{-3}$ & 3.05$\times 10^{-3}$ & -2.02$\times 10^{-2}$ &  & 2.18$\times 10^{-11}$ & 1.53$\times 10^{-3}$ & 2.08$\times 10^{-3}$ & 1.36$\times 10^{-3}$ \\
LAD            & 3.35$\times 10^{-6}$ & 3.55$\times 10^{-3}$ & 1.80$\times 10^{-4}$ & -1.76$\times 10^{-2}$ &  & 2.08$\times 10^{-11}$ & 1.74$\times 10^{-3}$ & 2.36$\times 10^{-3}$ & 1.32$\times 10^{-3}$ \\
Huber          & 3.53$\times 10^{-6}$ & 5.54$\times 10^{-3}$ & 4.37$\times 10^{-3}$ & -1.71$\times 10^{-2}$ &  & 2.16$\times 10^{-11}$ & 1.84$\times 10^{-3}$ & 2.53$\times 10^{-3}$ & 1.27$\times 10^{-3}$ \\
$\mu$-estimator & 2.84$\times 10^{-6}$ & 2.48$\times 10^{-3}$ & 1.16$\times 10^{-3}$ & -1.60$\times 10^{-2}$ &  & 1.91$\times 10^{-11}$ & 2.18$\times 10^{-3}$ & 3.06$\times 10^{-3}$ & 1.65$\times 10^{-3}$ \\
Cauchy         & 2.66$\times 10^{-6}$ & 1.60$\times 10^{-3}$ & 1.57$\times 10^{-3}$ & -1.55$\times 10^{-2}$ &  & 2.03$\times 10^{-11}$ & 2.51$\times 10^{-3}$ & 3.58$\times 10^{-3}$ & 1.94$\times 10^{-3}$ \\ \hline
\textbf{DE} &&&&&&&&&  \\
QMLE	&2.51$\times 10^{-6}$ &	1.42$\times 10^{-2}$&	-1.23$\times 10^{-2}$&	-1.77$\times 10^{-2}$ & & 1.49$\times 10^{-11}$&	2.59$\times 10^{-3}$&	2.59$\times 10^{-3}$&	1.35$\times 10^{-3}$\\
LAD	&1.74$\times 10^{-6}$&	  1.14$\times 10^{-2}$&	-1.09$\times 10^{-2}$&	-1.31$\times 10^{-2}$& & 6.60$\times 10^{-12}$&	  1.45$\times 10^{-3}$&	  1.84$\times 10^{-3}$&	  8.53$\times 10^{-4}$\\
Huber's	&1.73$\times 10^{-6}$&	1.21$\times 10^{-2}$&	-1.21$\times 10^{-2}$&	-1.28$\times 10^{-2}$ & & 6.73$\times 10^{-12}$	& 1.49$\times 10^{-3}$	& 1.92$\times 10^{-3}$	& 8.93$\times 10^{-4}$\\
$\mu$-estimator	&1.44$\times 10^{-6}$&	1.25$\times 10^{-2}$&	-7.18$\times 10^{-3}$&	-1.12$\times 10^{-2}$& &   5.64$\times 10^{-12}$	& 1.80$\times 10^{-3}$	& 2.46$\times 10^{-3}$	& 8.97$\times 10^{-4}$\\
Cauchy	&  1.37$\times 10^{-6}$&	1.36$\times 10^{-2}$&	  -5.67$\times 10^{-3}$&	  -1.12$\times 10^{-2}$ & & 6.61$\times 10^{-12}$	& 2.43$\times 10^{-3}$&	3.28$\times 10^{-3}$	&1.03$\times 10^{-3}$\\ \hline\hline

\textbf{Logistic} &&&&&&&&& \\
QMLE	&3.83$\times 10^{-6}$&	1.38$\times 10^{-2}$&	-1.73$\times 10^{-2}$&	-1.75$\times 10^{-2}$& & 2.64$\times 10^{-11}$ &	3.78$\times 10^{-3}$ &	3.01$\times 10^{-3}$ &	1.57$\times 10^{-3}$  \\
LAD	&2.97$\times 10^{-6}$&	8.27$\times 10^{-3}$&	-1.43$\times 10^{-2}$&	-1.20$\times 10^{-2}$& & 1.55$\times 10^{-11}$ &	2.01$\times 10^{-3}$	& 2.16$\times 10^{-3}$	&   1.11$\times 10^{-3}$\\
Huber's & 3.03$\times 10^{-6}$ &	8.42$\times 10^{-3}$ &	-1.23$\times 10^{-2}$ &	-1.25$\times 10^{-2}$& & 1.64$\times 10^{-11}$	&   2.01$\times 10^{-3}$ &	  2.03$\times 10^{-3}$	&1.12$\times 10^{-3}$  \\
$\mu$-estimator	& 2.50$\times 10^{-6}$ &	  6.28$\times 10^{-3}$ &	-1.25$\times 10^{-2}$&	-8.64$\times 10^{-3}$& &   1.33$\times 10^{-11}$	& 2.19$\times 10^{-3}$& 	2.98$\times 10^{-3}$&	1.23$\times 10^{-3}$\\
Cauchy	&  2.41$\times 10^{-6}$&	6.46$\times 10^{-3}$&	  -1.10$\times 10^{-2}$&	  -8.62$\times 10^{-3}$& & 1.42$\times 10^{-11}$ &	2.50$\times 10^{-3}$	& 3.49$\times 10^{-3}$	& 1.46$\times 10^{-3}$\\ \hline

${\boldsymbol {t(3)}}$         &          &          &           &           &  &          &          &          &          \\
QMLE           & 1.67$\times 10^{-6}$ & 2.89$\times 10^{-2}$ & -2.20$\times 10^{-2}$ & -3.48$\times 10^{-2}$ &  & 2.74$\times 10^{-11}$ & 1.37$\times 10^{-2}$ & 1.56$\times 10^{-2}$ & 8.02$\times 10^{-3}$ \\
LAD            & 1.00$\times 10^{-6}$ & 7.28$\times 10^{-3}$ & -6.13$\times 10^{-3}$ & -1.04$\times 10^{-2}$ &  & 5.62$\times 10^{-12}$ & 3.01$\times 10^{-3}$ & 4.58$\times 10^{-3}$ & 2.02$\times 10^{-3}$ \\
Huber's          & 9.74$\times 10^{-7}$ & 8.20$\times 10^{-3}$ & -8.00$\times 10^{-3}$ & -1.05$\times 10^{-2}$ &  & 5.50$\times 10^{-12}$ & 2.99$\times 10^{-3}$ & 4.53$\times 10^{-3}$ & 2.01$\times 10^{-3}$ \\
$\mu$-estimator & 6.62$\times 10^{-7}$ & 8.42$\times 10^{-3}$ & -8.91$\times 10^{-3}$ & -5.33$\times 10^{-3}$ &  & 3.93$\times 10^{-12}$ & 2.30$\times 10^{-3}$ & 3.59$\times 10^{-3}$ & 1.63$\times 10^{-3}$ \\
Cauchy         & 5.89$\times 10^{-7}$ & 9.44$\times 10^{-3}$ & -9.33$\times 10^{-3}$ & -5.20$\times 10^{-3}$ &  & 4.33$\times 10^{-12}$ & 2.51$\times 10^{-3}$ & 3.91$\times 10^{-3}$ & 1.85$\times 10^{-3}$ \\ \hline

$\boldsymbol{t(2.2)}$       &           &          &           &           &  &          &          &          &          \\
QMLE           & -4.35$\times 10^{-7}$ & 9.90$\times 10^{-2}$ & -4.39$\times 10^{-2}$ & -1.54$\times 10^{-1}$ &  & 1.90$\times 10^{-11}$ & 1.34$\times 10^{-1}$ & 1.48$\times 10^{-1}$ & 8.10$\times 10^{-2}$ \\
LAD            & 1.13$\times 10^{-6}$  & 3.16$\times 10^{-2}$ & -8.87$\times 10^{-5}$ & -3.48$\times 10^{-2}$ &  & 1.35$\times 10^{-11}$ & 3.30$\times 10^{-2}$ & 4.54$\times 10^{-2}$ & 1.38$\times 10^{-2}$ \\
Huber          & 1.38$\times 10^{-6}$  & 5.30$\times 10^{-2}$ & -1.08$\times 10^{-2}$ & -4.40$\times 10^{-2}$ &  & 1.53$\times 10^{-11}$ & 4.43$\times 10^{-2}$ & 5.52$\times 10^{-2}$ & 1.58$\times 10^{-2}$ \\
$\mu$-estimator & 4.55$\times 10^{-7}$  & 1.60$\times 10^{-2}$ & -4.41$\times 10^{-3}$ & -1.30$\times 10^{-2}$ &  & 5.51$\times 10^{-12}$ & 5.75$\times 10^{-3}$ & 9.33$\times 10^{-3}$ & 5.38$\times 10^{-3}$ \\
Cauchy         & 4.69$\times 10^{-7}$  & 2.04$\times 10^{-2}$ & -5.37$\times 10^{-3}$ & -1.47$\times 10^{-2}$ &  & 6.74$\times 10^{-12}$ & 6.13$\times 10^{-3}$ & 1.06$\times 10^{-2}$ & 6.52$\times 10^{-3}$ \\ \hline\hline
\end{tabular}
\end{table}

The adjusted biases and MSEs of  various M-estimators are reported in Table \ref{Sim.GARCH21}. It turns out that  the bias and MSE of all M-estimators are quite close to those of the QMLE under normal errors. However, the QMLE produces biases and MSEs that are sizeably larger than those for the other M-estimators under heavier tail distributions. Under the $t(3)$ and $t(2.2)$ distributions with infinite fourth-order moments, the advantage of the M-estimators over the QMLE becomes more prominent. Also, under the $t(2.2)$ distribution, the LAD and Huber estimators perform poorly compared with the $\mu$- and Cauchy-estimators since the former two yield significantly larger MSE than the latter two. This provides some evidence to support the following: 
\begin{enumerate}
\item[(i)] under Gaussian error distributions, all M-estimators have similar performance;
\item[(ii)] the better performance of some M-estimators under heavy-tail error distributions does not come at the cost of a loss of efficiency under normal error distribution, and 
\item[(iii)] the $\mu$- and Cauchy- M-estimators are less sensitive to the heavy-tail errors than the LAD and Huber estimators.
\end{enumerate}

\subsection{A misspecified GARCH case}\label{missp}

It is of interest to check whether the M-estimators remain consistent when the order of a GARCH model is misspecified. In particular, we consider overfitting a GARCH $(p_0,q_0)$ with a higher-order GARCH $(p,q)$ model when at least
one of $p>p_0$ or $q>q_0$ holds. In this case, we are essentially fitting a GARCH model with some component(s) of the parameter~$\bth$ equal to zero (hence lying on the boundary of the parameter space, a case which is not covered by the consistency results available so far). However, a numerical exploration of a GARCH (1,1)  misspecified as GARCH (2,1) indicates that consistency can be expected to hold under such overfitting as provided below.  

Various M-estimators of a GARCH~(2,1) were computed from  simulated GARCH~(1,1) series with  parameter  value~$\bth_0=(1.65\times 10^{-5}, 0.0701, 0.901)^\prime$ and various error distributions (sample size $n =1000$ and~$R = 1000$ replications). 
 The adjusted bias and MSE of the M-estimators are shown in Table \ref{misspecified.Sim} by wrongly fitting  a GARCH~(2,1) with  parameter value~$(1.65\times 10^{-5}, 0.0701, 0, 0.901)^\prime$. For all distributions considered, the M-estimates of the spurious parameter $\alpha_2$ is close to zero, and the bias  and the MSE are  quite small, indicating good performance of the M-estimators despite the misspecification. As in Table \ref{Sim.GARCH21}, however, the QMLE appears to be sensitive to the heavy-tailed distributions while other M-estimators are more robust.

\begin{table}[!htbp]
\caption{The adjusted bias and MSE of the M-estimators under a GARCH~(1,1) model  misspecified  as GARCH~(2,1) under various error distributions (Normal, double exponential, logistic, $t(3)$); sample size~$n =~\!1000$; $R = 1000$ replications.\vspace{3mm}}\label{misspecified.Sim}
\centering
\scriptsize  
\begin{tabular}{c c c c c c c c c c}\hline\hline 
&\multicolumn{4}{c}{} &&\multicolumn{4}{c}{ \vspace{-2mm}}\\
&\multicolumn{4}{c}{\textbf{adjusted bias}} &&\multicolumn{4}{c}{\textbf{adjusted MSE}\vspace{1.5mm}} \\ \cline{2-5} \cline{7-10} & $\omega$ &$\alpha_1$&$\alpha_2(=0)$&$\beta$&&$\omega$ &$\alpha_1$&$\alpha_2(=0)$&$\beta$ \\ \hline
\textbf{Normal}&&&&&&& \\
QMLE     & 1.11$\times 10^{-5}$ & -2.00$\times 10^{-3}$ & 5.97$\times 10^{-3}$ & {  -2.38$\times 10^{-2}$}
 && 3.94$\times 10^{-10}$ &   1.55$\times 10^{-3}$ &  1.87$\times 10^{-3}$ &   2.64$\times 10^{-3}$\\
LAD &    1.09$\times 10^{-5}$ &  -1.73$\times 10^{-3}$ &  5.65$\times 10^{-3}$ & -2.43$\times 10^{-2}$
 &&4.53$\times 10^{-10}$ & 1.73$\times 10^{-3}$ & 2.12$\times 10^{-3}$ & 3.09$\times 10^{-3}$\\
Huber's &  1.22$\times 10^{-5}$ &  1.25$\times 10^{-3}$ & 6.08$\times 10^{-3}$ & -2.43$\times 10^{-2}$
 && 5.18$\times 10^{-10}$ & 1.82$\times 10^{-3}$ & 2.28$\times 10^{-3}$ & 3.13$\times 10^{-3}$ \\
$\mu$-estimator &1.11$\times 10^{-5}$ &   -5.36$\times 10^{-4}$ &  5.75$\times 10^{-3}$ & -2.49$\times 10^{-2}$
&& 5.27$\times 10^{-10}$ & 2.42$\times 10^{-3}$ & 2.99$\times 10^{-3}$ & 3.67$\times 10^{-3}$\\
Cauchy &1.13$\times 10^{-5}$ & -5.85$\times 10^{-4}$ & 6.31$\times 10^{-3}$ & -2.61$\times 10^{-2}$
  && 6.26$\times 10^{-10}$ & 2.83$\times 10^{-3}$ & 3.57$\times 10^{-3}$ & 4.41$\times 10^{-3}$ \\ \hline  
\textbf{DE} &&&&&&&\\
QMLE     & 9.70$\times 10^{-6}$ & -1.07$\times 10^{-3}$ & 7.12$\times 10^{-3}$ & -2.45$\times 10^{-2}$
&&4.19$\times 10^{-10}$ & 2.82$\times 10^{-3}$ & 3.33$\times 10^{-3}$ & 3.78$\times 10^{-3}$\\
LAD & 8.11$\times 10^{-6}$ &   6.07$\times 10^{-4}$ &  4.72$\times 10^{-3}$ & -1.89$\times 10^{-2}$
&&2.91$\times 10^{-10}$& 2.24$\times 10^{-3}$ &2.60$\times 10^{-3}$ &  2.51$\times 10^{-3}$ \\
Huber's & 7.84$\times 10^{-6}$ & -7.00$\times 10^{-4}$ & 4.79$\times 10^{-3}$ & -1.94$\times 10^{-2}$
&&2.92$\times 10^{-10}$ &   2.20$\times 10^{-3}$ &    2.54$\times 10^{-3}$ & 2.58$\times 10^{-3}$\\
$\mu$-estimator &   7.21$\times 10^{-6}$ & 2.45$\times 10^{-3}$ &   3.15$\times 10^{-3}$ &   -1.69$\times 10^{-2}$
&&  2.85$\times 10^{-10}$ & 2.59$\times 10^{-3}$& 3.02$\times 10^{-3}$& 2.59$\times 10^{-3}$ \\
Cauchy &7.49$\times 10^{-6}$ &  3.86$\times 10^{-3}$ &  3.29$\times 10^{-3}$ &-1.79$\times 10^{-2}$
&&3.48$\times 10^{-10}$ & 3.10$\times 10^{-3}$ & 3.65$\times 10^{-3}$ & 3.20$\times 10^{-3}$  \\ \hline
 
\textbf{Logistic}       &&&&&&&\\
QMLE     & 1.24$\times 10^{-5}$ & -1.95$\times 10^{-3}$ & 9.70$\times 10^{-3}$& -2.68$\times 10^{-2}$
 && 5.24$\times 10^{-10}$ & 2.14$\times 10^{-3}$&  2.61$\times 10^{-3}$ & 3.28$\times 10^{-3}$\\
LAD &1.03$\times 10^{-5}$& -2.81$\times 10^{-3}$& 8.40$\times 10^{-3}$& -2.30$\times 10^{-2}$&&
3.88$\times 10^{-10}$ & 1.82$\times 10^{-3}$ & 2.23$\times 10^{-3}$ &2.63$\times 10^{-3}$ \\
Huber's &1.00$\times 10^{-5}$&	-3.27$\times 10^{-3}$ &  8.11$\times 10^{-3}$ & -2.28$\times 10^{-2}$
&&  3.83$\times 10^{-10}$ &   1.78$\times 10^{-3}$ &  2.14$\times 10^{-3}$ &   2.62$\times 10^{-3}$  \\
$\mu$-estimator &  9.47$\times 10^{-6}$ & -2.29$\times 10^{-3}$ & 8.31$\times 10^{-3}$ &   -2.21$\times 10^{-2}$
&&3.88$\times 10^{-10}$ & 2.15$\times 10^{-3}$& 2.69$\times 10^{-3}$ & 2.86$\times 10^{-3}$ \\
Cauchy & 9.74$\times 10^{-6}$&   -8.90$\times 10^{-4}$ & 8.56$\times 10^{-3}$ & -2.26$\times 10^{-2}$
&& 4.34$\times 10^{-10}$ & 2.53$\times 10^{-3}$ & 3.21$\times 10^{-3}$ & 3.23$\times 10^{-3}$ \\ \hline  
 
${\boldsymbol {t(3)}}$ & &&&&&&&\\
QMLE     & 1.08$\times 10^{-5}$ & 1.64$\times 10^{-2}$  & 1.14$\times 10^{-2}$ & -5.47$\times 10^{-2}$
 &&1.15$\times 10^{-9}$ & 1.93$\times 10^{-2}$& 2.67$\times 10^{-2}$ & 1.97$\times 10^{-2}$\\
LAD & 4.50$\times 10^{-6}$ &   1.05$\times 10^{-3}$ & 2.96$\times 10^{-3}$ &-2.08$\times 10^{-2}$
 &&1.85$\times 10^{-10}$ & 3.01$\times 10^{-3}$ & 3.41$\times 10^{-3}$ & 3.39$\times 10^{-3}$\\
Huber's &  5.46$\times 10^{-6}$ & 4.83$\times 10^{-3}$ & 2.64$\times 10^{-3}$ & -2.03$\times 10^{-2}$
 &&2.19$\times 10^{-10}$& 3.33$\times 10^{-3}$ & 3.80$\times 10^{-3}$& 3.50$\times 10^{-3}$ \\
 $\mu$-estimator &  4.47$\times 10^{-6}$  & 5.91$\times 10^{-3}$  & 4.41$\times 10^{-4}$ &   -1.51$\times 10^{-2}$
 &&   1.45$\times 10^{-10}$ & 2.55$\times 10^{-3}$ &  2.84$\times 10^{-3}$ &   2.25$\times 10^{-3}$\\
Cauchy &    3.65$\times 10^{-6}$ & 3.85$\times 10^{-3}$ &   4.77$\times 10^{-5}$ & -1.54$\times 10^{-2}$
&& 1.45$\times 10^{-10}$ &   2.51$\times 10^{-3}$ & 2.86$\times 10^{-3}$ & 2.56$\times 10^{-3}$
\\ \hline\hline  
\end{tabular}
\end{table} 


\subsection{GARCH~(1,2) models}\label{GARCH12}
Simulations for the GARCH~(1,2) (with parameter $\boldsymbol \theta_0=(0.1, 0.1, 0.2, 0.6)^\prime$, $R=1000$, and~$n=1000$)  were conducted in the same way as for GARCH~(2,1) in Section~\ref{GARCH21}. The results are shown in Table \ref{Sim.GARCH12};  we do not report the results for the QMLE 
 under the $t(3)$ and~$t(2.2)$ error distributions, since the algorithm did not converge for most replications; a failure of the QMLE. 

Inspection of Table \ref{Sim.GARCH12} reveals that under normal error distribution, the LAD and Huber estimators produce MSEs that are close to the QMLE ones while the $\mu$- and Cauchy M-estimators yield larger MSEs   for the estimation of~$\omega$ and~$\alpha$. For the double exponential  and logistic distributions, there is no significant difference between the various estimators. Clear difference emerges under heavy-tailed distributions though; 
the $\mu$- and Cauchy M-estimators produce smaller MSEs than the LAD and Huber estimators of $\omega$  
and~$\alpha$ under the~$t(3)$ and~$t(2.2)$ distributions, respectively. 

\begin{table}[!htbp]
\caption{ The adjusted bias and MSE of   M-estimators for GARCH~(1, 2) models under various error distributions (Normal, double exponential, logistic, $t(3)$, $t(2.2)$); sample size~$n =~\!1000$; $R = 1000$ replications.\vspace{3mm}}\label{Sim.GARCH12}
\centering
\scriptsize  
\begin{tabular}{c c c c c c c c c c}\hline\hline 
&\multicolumn{4}{c}{\textbf{}} &&\multicolumn{4}{c}{\textbf{}\vspace{-2mm}} \\ 
&\multicolumn{4}{c}{\textbf{adjusted bias}} &&\multicolumn{4}{c}{\textbf{adjusted MSE}\vspace{1.5mm}} \\ \cline{2-5} \cline{7-10} & $\omega$ &$\alpha$& $\beta_1$& $\beta_2$&  &$\omega$ &$\alpha$ & $\beta_1$&$\beta_2$ \\ \hline
Normal            &          &           &           &           &                          &          &            &            &          \\
QMLE              & 5.53$\times 10^{-2}$ & 1.10$\times 10^{-3}$  & 9.65$\times 10^{-2}$  & -1.52$\times 10^{-1}$ &                          & 2.66$\times 10^{-2}$ & 1.17$\times 10^{-3}$   & 1.45$\times 10^{-1}$   & 1.38$\times 10^{-1}$ \\
LAD               & 5.93$\times 10^{-2}$ & 7.15$\times 10^{-4}$  & 9.01$\times 10^{-2}$  & -1.50$\times 10^{-1}$ &                          & 3.21$\times 10^{-2}$ & 1.31$\times 10^{-3}$   & 1.55$\times 10^{-1}$   & 1.45$\times 10^{-1}$ \\
Huber             & 6.49$\times 10^{-2}$ & 4.64$\times 10^{-3}$  & 9.77$\times 10^{-2}$  & -1.57$\times 10^{-1}$ &                          & 3.72$\times 10^{-2}$ & 1.37$\times 10^{-3}$   & 1.56$\times 10^{-1}$   & 1.47$\times 10^{-1}$ \\
$\mu$-estimator   & 7.45$\times 10^{-2}$ & 8.93$\times 10^{-4}$  & 1.11$\times 10^{-1}$  & -1.86$\times 10^{-1}$ &                          & 7.41$\times 10^{-2}$ & 1.84$\times 10^{-3}$   & 2.16$\times 10^{-1}$   & 2.01$\times 10^{-1}$ \\
Cauchy            & 7.51$\times 10^{-2}$ & 1.25$\times 10^{-3}$  & 1.29$\times 10^{-1}$  & -2.06$\times 10^{-1}$ &                          & 6.30$\times 10^{-2}$ & 2.17$\times 10^{-3}$   & 2.43$\times 10^{-1}$   & 2.31$\times 10^{-1}$ \\ \hline
DE                &          &           &           &           &                          &          &            &            &          \\
QMLE              & 5.48$\times 10^{-2}$ & 2.93$\times 10^{-3}$  & 1.01$\times 10^{-1}$  & -1.63$\times 10^{-1}$ &                          & 3.15$\times 10^{-2}$ & 1.79$\times 10^{-3}$   & 1.62$\times 10^{-1}$   & 1.57$\times 10^{-1}$ \\
LAD               & 3.73$\times 10^{-2}$ & -1.93$\times 10^{-3}$ & 8.76$\times 10^{-2}$  & -1.27$\times 10^{-1}$ &                          & 1.20$\times 10^{-2}$ & 1.61$\times 10^{-3}$   & 1.46$\times 10^{-1}$   & 1.35$\times 10^{-1}$ \\
Huber             & 3.83$\times 10^{-2}$ & -1.22$\times 10^{-3}$ & 9.51$\times 10^{-2}$  & -1.36$\times 10^{-1}$ &                          & 1.21$\times 10^{-2}$ & 1.65$\times 10^{-3}$   & 1.53$\times 10^{-1}$   & 1.44$\times 10^{-1}$ \\
$\mu$-estimator   & 4.05$\times 10^{-2}$ & 1.15$\times 10^{-3}$  & 1.13$\times 10^{-1}$  & -1.52$\times 10^{-1}$ &                          & 1.72$\times 10^{-2}$ & 2.05$\times 10^{-3}$   & 1.73$\times 10^{-1}$   & 1.60$\times 10^{-1}$ \\
Cauchy            & 4.74$\times 10^{-2}$ & 3.26$\times 10^{-3}$  & 1.18$\times 10^{-1}$  & -1.66$\times 10^{-1}$ &                          & 2.55$\times 10^{-2}$ & 2.48$\times 10^{-3}$   & 1.85$\times 10^{-1}$   & 1.72$\times 10^{-1}$ \\ \hline
Logistic          &          &           &           &           &                          &          &            &            &          \\
QMLE              & 5.77$\times 10^{-2}$ & 2.76$\times 10^{-3}$  & 1.06$\times 10^{-1}$  & -1.61$\times 10^{-1}$ &                          & 3.02$\times 10^{-2}$ & 1.49$\times 10^{-3}$   & 1.67$\times 10^{-1}$   & 1.59$\times 10^{-1}$ \\
LAD               & 4.50$\times 10^{-2}$ & -5.78$\times 10^{-5}$ & 7.27$\times 10^{-2}$  & -1.18$\times 10^{-1}$ &                          & 1.58$\times 10^{-2}$ & 1.37$\times 10^{-3}$   & 1.30$\times 10^{-1}$   & 1.18$\times 10^{-1}$ \\
Huber             & 4.50$\times 10^{-2}$ & -2.33$\times 10^{-4}$ & 8.85$\times 10^{-2}$  & -1.34$\times 10^{-1}$ &                          & 1.58$\times 10^{-2}$ & 1.36$\times 10^{-3}$   & 1.53$\times 10^{-1}$   & 1.39$\times 10^{-1}$ \\
$\mu$-estimator   & 4.52$\times 10^{-2}$ & 1.32$\times 10^{-3}$  & 9.39$\times 10^{-2}$  & -1.40$\times 10^{-1}$ &                          & 1.80$\times 10^{-2}$ & 1.72$\times 10^{-3}$   & 1.58$\times 10^{-1}$   & 1.44$\times 10^{-1}$ \\
Cauchy            & 5.15$\times 10^{-2}$ & 2.91$\times 10^{-3}$  & 1.05$\times 10^{-1}$  & -1.57$\times 10^{-1}$ &                          & 2.98$\times 10^{-2}$ & 2.08$\times 10^{-3}$   & 1.85$\times 10^{-1}$   & 1.70$\times 10^{-1}$ \\ \hline
$\boldsymbol{t(3)}$            &          &           &           &           &                          &          &            &            &          \\
QMLE              &     -     &     -      &       -    &  -         &                          &      -    &     -       &      -      &       -   \\
LAD               & 2.93$\times 10^{-2}$ & 2.43$\times 10^{-3}$  & 1.08$\times 10^{-1}$  & -1.40$\times 10^{-1}$ &                          & 1.13$\times 10^{-2}$ & 2.49$\times 10^{-3}$   & 1.82$\times 10^{-1}$   & 1.59$\times 10^{-1}$ \\
Huber             & 2.87$\times 10^{-2}$ & 1.50$\times 10^{-3}$  & 9.13$\times 10^{-2}$  & -1.26$\times 10^{-1}$ &                          & 1.18$\times 10^{-2}$ & 2.30$\times 10^{-3}$   & 1.60$\times 10^{-1}$   & 1.40$\times 10^{-1}$ \\
$\mu$-estimator   & 1.57$\times 10^{-2}$ & 8.75$\times 10^{-5}$  & 1.21$\times 10^{-1}$  & -1.37$\times 10^{-1}$ &                          & 5.59$\times 10^{-3}$ & 1.88$\times 10^{-3}$   & 1.63$\times 10^{-1}$   & 1.42$\times 10^{-1}$ \\
Cauchy            & 1.50$\times 10^{-2}$ & 6.44$\times 10^{-4}$  & 1.38$\times 10^{-1}$  & -1.54$\times 10^{-1}$ &                          & 6.50$\times 10^{-3}$ & 2.15$\times 10^{-3}$   & 1.90$\times 10^{-1}$   & 1.65$\times 10^{-1}$ \\ \hline
$\boldsymbol{t(2.2)}$          &          &           &           &           &                          &          &            &            &          \\
QMLE              &     -     &   -        &        -   &     -      &                          &       -   &     -       &      -      &    -      \\
LAD               & 3.53$\times 10^{-2}$ & 2.57$\times 10^{-2}$  & 1.24$\times 10^{-1}$  & -1.85$\times 10^{-1}$ &                          & 1.30$\times 10^{-2}$ & 1.41$\times 10^{-2}$   & 2.41$\times 10^{-1}$   & 2.21$\times 10^{-1}$ \\
Huber             & 4.86$\times 10^{-2}$ & 3.99$\times 10^{-2}$  & 7.81$\times 10^{-2}$  & -1.66$\times 10^{-1}$ &                          & 1.44$\times 10^{-2}$ & 1.63$\times 10^{-2}$   & 1.81$\times 10^{-1}$   & 1.79$\times 10^{-1}$ \\
$\mu$-estimator   & 1.72$\times 10^{-2}$ & 5.18$\times 10^{-3}$  & 1.51$\times 10^{-1}$  & -1.78$\times 10^{-1}$ &                          & 1.73$\times 10^{-2}$ & 4.27$\times 10^{-3}$   & 2.42$\times 10^{-1}$   & 2.12$\times 10^{-1}$ \\
Cauchy            & 2.15$\times 10^{-2}$ & 9.68$\times 10^{-3}$  & 1.50$\times 10^{-1}$  & -1.85$\times 10^{-1}$ &                          & 2.05$\times 10^{-2}$ & 4.90$\times 10^{-3}$   & 2.34$\times 10^{-1}$   & 2.14$\times 10^{-1}$ \\ \hline\hline
\end{tabular}
\end{table}

\begin{table}[!htbp]\thisfloatpagestyle{empty}
\caption{The coverage rates (in percentage) of the bootstrap schemes M, E and U and asymptotic normal approximations for the M-estimators QMLE, LAD, Huber's, $\mu$- and Cauchy-; the error distributions are 
normal and $t(3)$.}\label{boot.cover}
\centering
\begin{tabular}{c c c c c c c c c c} \hline\hline
&&&\multicolumn{3}{c}{$90\%$ nominal level} &&\multicolumn{3}{c}{$95\%$ nominal level} \\ \cline{4-6} \cline{8-10} & &&$\omega$ &$\alpha$&$\beta$ & & $\omega$ &$\alpha$&$\beta$ \\ \hline
Normal& QMLE & Scheme M & 89.0 &	86.2 	&88.2 	&&	91.0 &	92.2 	&91.4  \\ 
&&Scheme E &  87.2 	&83.8 &86.8 	&&	90.2 &	88.4 	&91.2  \\
&&Scheme U &  90.2 &	87.4 	&87.2 	&&	94.4 &	92.6 	&93.2  \\
&&Asymptotic & 82.6  &	91.0  & 85.8  &&	87.0 &	95.2  &	89.0  \\ \hline
Normal& LAD&Scheme M & 86.0 	&83.4 &	84.2 &&	88.2 	&87.2 	&88.4 \\
&&Scheme E & 88.0 	&87.2 	&87.2 	&&	91.0 	& 91.2 	& 90.2  \\
&&Scheme U & 88.6 	&88.4 	&88.0 	&&	93.2 	& 91.8 	& 91.8  \\
&&Asymptotic & 94.0 	& 98.8 &	87.0  &&	96.4  & 99.4  &	90.4  \\ \hline
Normal & Huber's&Scheme M & 88.8 	& 85.4 & 86.6 &&	91.2 &	89.8 	& 91.2  \\
&&Scheme E & 88.2 	& 89.0 & 88.0  &&	91.4 	& 92.4  &	90.0   \\
&&Scheme U & 89.6 	& 90.4 & 88.4  &&	93.6 	& 93.6  &	91.8  \\ 
&&Asymptotic & 87.6 &	95.4 &	86.2 &&		90.6 &	96.6 &	90.4 
\\ \hline
Normal & $\mu$-estimator &Scheme M & 88.0 	&84.6 	&86.8 	&&	89.6 	&87.8 	& 88.6  \\
&&Scheme E & 87.4 	&84.8 	&86.6 		&&89.4 	&88.4 	& 88.4  \\
&&Scheme U & 88.6 	& 88.4 &87.6 	&&	91.8 	&91.8 	&90.6  \\ 
&&Asymptotic & 71.4 	&69.6  &	86.8  &&	77.4 	&78.2 &	90.8 \\ \hline
Normal &Cauchy& Scheme M & 85.6 	&84.0 	&84.4 	&&	87.8 	&85.8 	& 87.6  \\
&&Scheme E & 81.4 	& 82.2 &80.2 	&&	82.8 	&86.2 	&84.2  \\
&&Scheme U & 88.4 	&88.2 	&87.0 	&&	90.4 	&91.4 	&89.4  \\ 
&&Asymptotic & 97.8  &	99.8  &	85.0 &&		98.2 &	100.0 &	89.6 \\ \hline
$t(3)$& QMLE &Scheme M & 71.0 	&75.4 	&74.8 	&&	75.0 &	79.0 	&78.0  \\
&&Scheme E &  67.6 & 72.4 	&66.8 	&&	73.4 	&76.2 	&72.4 \\ 
&&Scheme U &  75.6  &84.6 & 75.0  &&	81.6 &	87.2 	&80.0  \\ 
&&Asymptotic & - & - & - && - & - & - \\ \hline
$t(3)$& LAD&Scheme M & 84.4 	&80.6 	&83.0 	&&	85.4 	&83.8 &	87.8  \\
&&Scheme E & 84.6 &	85.0 &	81.4 	&&	87.6 &	87.0 &	86.6 \\
&&Scheme U & 81.6 	&86.2 & 79.2 &&	87.4 &	89.2 &	84.8 \\
&&Asymptotic & 98.0 &	99.8 &	88.8 &&		99.6 &	100.0 &	91.2 \\ \hline
$t(3)$& Huber's &Scheme M & 83.0 &	80.6 &	81.8 &&	85.6 &	83.2 &	86.6  \\
&&Scheme E & 81.8 &	79.2 &	80.8 &&	85.8 &	81.6 &	85.8 \\
&&Scheme U & 86.2 &88.0 &	86.0 &&	90.2 &	91.4 &	90.2  \\ 
&&Asymptotic & 96.8  &	99.0 &	88.4  &&	97.8  &	99.6 &	92.8 
\\ \hline
$t(3)$&$\mu$-estimator&Scheme M & 82.4 &	84.8 &	83.8 &&		86.2 &	88.4 &	88.2 \\
&&Scheme E & 84.6 &	84.0 &	84.6 &&	87.4 &	88.0 &	88.8 \\
&&Scheme U & 82.6 &	83.6 &	80.4 && 88.8 &	88.2 &	86.4  \\ 
&&Asymptotic & 86.6 &	91.8 &	80.8 &&	90.6 &	95.6 &	86.4  \\   \hline
$t(3)$&Cauchy &Scheme M & 78.2 &	83.4 &	78.4 &&	81.8 &	86.2 &	82.0 \\
&&Scheme E & 83.4 &	85.6 &	82.6 &&	85.4 &	89.0 &	87.2 \\
&&Scheme U & 85.0 &	85.0 &	84.8 &&	90.0 &	88.6 &	89.2 \\ 
&&Asymptotic & 100.0 	&100.0 &	85.6 &&		100.0 &	100.0 &	90.8  \\ \hline\hline
\end{tabular}
\end{table}

\section{Performance of the bootstrap confidence intervals }\label{sec.simBootMest}
The  performance of bootstrap based on various bootstrap schemes and classical confidence intervals  (based on the QMLE) can be assessed and compared in terms of coverage rates. We generated $R=500$ series of length $n=1000$ from the GARCH $(1, 1)$ model with parameter 
value~$\bth_0 = (0.1, 0.1, 0.8)^\prime$, under  normal and $t(3)$ error distributions. For each simulated series, we computed $B=2000$ bootstrap estimates based on the bootstrap schemes M, E, and~U described in Section~\ref{sec.BootMest} and constructed the bootstrap  and asymptotic confidence intervals   using~(\ref{bci}) and~(\ref{nci}), respectively. The coverage rates are computed as the proportions of the confidence intervals   covering the actual parameter value. In Table~\ref{boot.cover}, 
we report these coverage rates (in percentage) for   nominal confidence levels $90\%$ and $95\%$.

Under the normal distribution, the coverage rates of the bootstrap approximation are generally close to the nominal levels. Also, the bootstrap approximation works better for the~QMLE, LAD, and Huber estimators than for the $\mu$- and Cauchy ones. However, under the~$t(3)$ distribution, the bootstrap approximation works poorly for the QMLE while the coverage rates are reasonably good for all other M-estimators. For both distributions, Scheme U outperforms Schemes M and E.  
Except  for the Gaussian case, thus, in terms of coverage rates, the classical confidence intervals based on the asymptotics of the QMLE are  outperformed by the bootstrap confidence intervals based on the bootstrap Scheme U
and is recommended in the analysis of the financial data. 

\section{Real data analysis}\label{sec.MestData}
In this section, we analyse   two financial   series of daily log-returns,   the FTSE 100 Index data from January 2007 to December 2009 ($n = 783$) and 
 the Electric Fuel Corporation (EFCX) data from January 2000 to December 2001 ($n = 498$).
 Based on exploratory data analysis, a~GARCH~(1, 1) model has been selected for the 
 EFCX. A GARCH~(2, 1) model was preferred
  for the FTSE 100 data for two reasons. First, when fitted by the GARCH~(2,~1) model (via the~{\tt fGarch} package in R), the parameter $\alpha_2$, with p-value   $0.019$, is highly significant; second, the Akaike information criterion (AIC) for the GARCH~(2,~\!1) model is smaller than that for the GARCH~(1, 1) model.

\subsection{ The FTSE 100 data 
}\label{FTSE100}
Table \ref{FTSE.Mest} shows the the estimates given by {\tt fGarch} and by our M-estimators when fitting a GARCH~(2, 1) model to the FTSE 100 data.   The QMLE (based on 
 (\ref{just3})) and {\tt fGarch} provide similar results for all components of the parameter. Also, the M-estimates of $\beta$ do not vary much. For $\omega$, $\alpha_1$, and $\alpha_2$, the M-estimates are quite different since $c_H$ in (\ref{thetanoth}) depends on the score function $H$ used for the estimation.

\begin{table}
\caption{FTSE 100 data. The M-estimates  (QMLE, LAD, Huber's, $\mu$- and Cauchy-) of the GARCH~(2, 1) model using the FTSE 100 data; the QMLEs are obtained by using {\tt fGarch} and (\ref{just3}).\medskip}\label{FTSE.Mest}
\centering
\begin{tabular}{c c c c c c c}
\hline\hline
  & {\tt fGarch} & QMLE & LAD & Huber's & $\mu$-estimator & Cauchy \\
\hline
$\omega$ & 4.46$\times 10^{-6}$ & 4.65$\times 10^{-6}$ & 3.13$\times 10^{-6}$ & 3.55$\times 10^{-6}$ & 1.02$\times 10^{-5}$ & 2.51$\times 10^{-6}$ \\
\hline
$\alpha_1$ &  5.25$\times 10^{-2}$ & 4.51$\times 10^{-2}$ & 2.46$\times 10^{-2}$ & 3.45$\times 10^{-2}$ & 4.95$\times 10^{-2}$ & 6.83$\times 10^{-3}$ \\
\hline
$\alpha_2$ &  0.11 & 9.00$\times 10^{-2}$ & 5.57$\times 10^{-2}$ & 6.42$\times 10^{-2}$ & 0.17 & 4.18$\times 10^{-2}$ \\
\hline
$\beta$ &  0.83 & 0.85 & 0.84 & 0.86 & 0.81 & 0.80 \\
\hline\hline
\end{tabular}
\end{table}

For a GARCH~($p, q$) model, using (\ref{vt.hat.cj}) and the formulas for $\{c_j(\bth); j \geq 0 \}$ in Section 3 of Berkes et al. (2003), we have $\hat{v}_t(\bth_{0H}) = c_H\hat{v}_t(\bth_0)$. Since an M-estimator $\hat{\bth}_n$ is an estimator of~$\bth_{0H}$, $\hat{v}_t(\hat{\bth}_n)$ estimates $c_H v_t(\bth_0)$,
which is a scale-transformation of the conditional variance. 
To examine the behavior of the market volatility after eliminating the effect of any particular M-estimator used, we define the 
normalized volatilities as 
\begin{equation}
\hat{u}_t(\hat{\bth}_n) \coloneqq  \hat{v}_t(\hat{\bth}_n)/\sum_{i=1}^n\hat{v}_i(\hat{\bth}_n); \,\, 1 \le t \le n.
\end{equation}

\begin{figure}[h!]
\caption{FTSE 100 data. The plot of the squared returns and the estimated normalized conditional variances using various M-estimators for the FTSE 100 data.\smallskip}
\centering
\includegraphics[trim=0 10mm 0 13mm, clip, scale=.7]{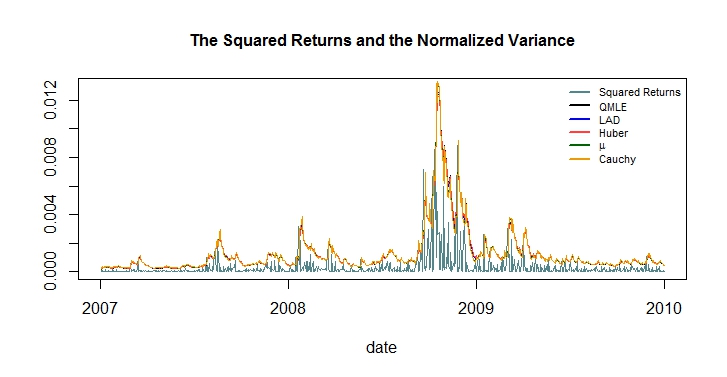}

\label{fig:variances}
\end{figure}

Figure \ref{fig:variances} shows the plot of $\{\hat{u}_t(\hat{\bth}_n); 1 \leq t \leq n \}$ based on various M-estimators against the squared returns. Notice that although the M-estimates in Table \ref{FTSE.Mest} are distinct, the plot of their normalized volatilities in Figure \ref{fig:variances} almost overlap. 
 Also, large values of the normalized volatilities and large squared returns occur at the same time. In this sense, the volatilities are well-modelled by the resulting GARCH(2, 1).

\subsection{ The Electric Fuel Corporation (EFCX) data }

 Fitting a GARCH~(1, 1) model to the EFCX data, Muler and Yohai (2008)  note that the QMLE and LAD estimates of the parameter
$\beta$ are significantly different. In Table~\ref{EFCX.Mest}, we report estimates given by the {\tt fGarch} and  
M-estimators. Note that in our previous analysis of the 
 FTSE 100 data, {\tt fGarch} estimates and the  QMLE were quite close, while their differences, for this EFCX data,  are much more prominent. It is also worth noting that while the LAD, Huber, $\mu-$ and Cauchy-estimates of $\beta$ are close to each other, they are all quite different from the corresponding value $0.84$ of the QMLE. 
That difference might be related to the infinite fourth moment of the underlying innovation distribution and the non-robustness of the QMLE. 

\begin{table}[!htbp]
\caption{EFCX data. The M-estimates (QMLE, LAD, Huber, $\mu$- and Cauchy) of the GARCH~(1, 1) model for the EFCX data; the QMLEs are obtained by using {\tt fGarch} and~(\ref{just3}).\vspace{3mm}}\label{EFCX.Mest}
\centering
\begin{tabular}{c c c c c c c}
\hline\hline 
  & {\tt fGarch} & QMLE & LAD & Huber's & $\mu$-estimator & Cauchy \\
\hline
$\omega$ & 1.89$\times 10^{-4}$	& 6.28$\times 10^{-4}$ & 6.43$\times 10^{-4}$	&8.37$\times 10^{-4}$&	1.42$\times 10^{-3}$&	2.97$\times 10^{-4}$ \\
\hline
$\alpha$ &  4.54$\times 10^{-2}$&7.20$\times 10^{-2}$&8.87$\times 10^{-2}$&	0.10 &	0.27 &	6.35$\times 10^{-2}$ \\
\hline
$\beta$ &  0.92 & 0.84 &	0.66 &	0.67 &	0.61 &	0.60 \\
\hline\hline 
\end{tabular}
\end{table}

To determine whether the innovation distribution may have finite fourth moment, we examine  QQ-plots of the residuals  
$\{X_t/ \hat{v}_t^{1/2}(\hat{\bth}_n); 1 \le t \le n\}$ based on the $\mu$-estimator $\hat{\bth}_n$ against Student  $t(d)$ distributions for various degrees of freedom $d$. We consider the $\mu$-estimator, which requires the mildest moment assumptions on the innovation distribution. 

\begin{figure}[!h]
\caption{EFCX and FTSE 100 data. The QQ-plot of the residuals against $t(d)$ distributions for the EFCX (left column, $d=4.01$ and 3.01) and FTSE 100 (right column, $d=4.01$ and~12.01) data.}
\centering
\includegraphics[scale=0.55]{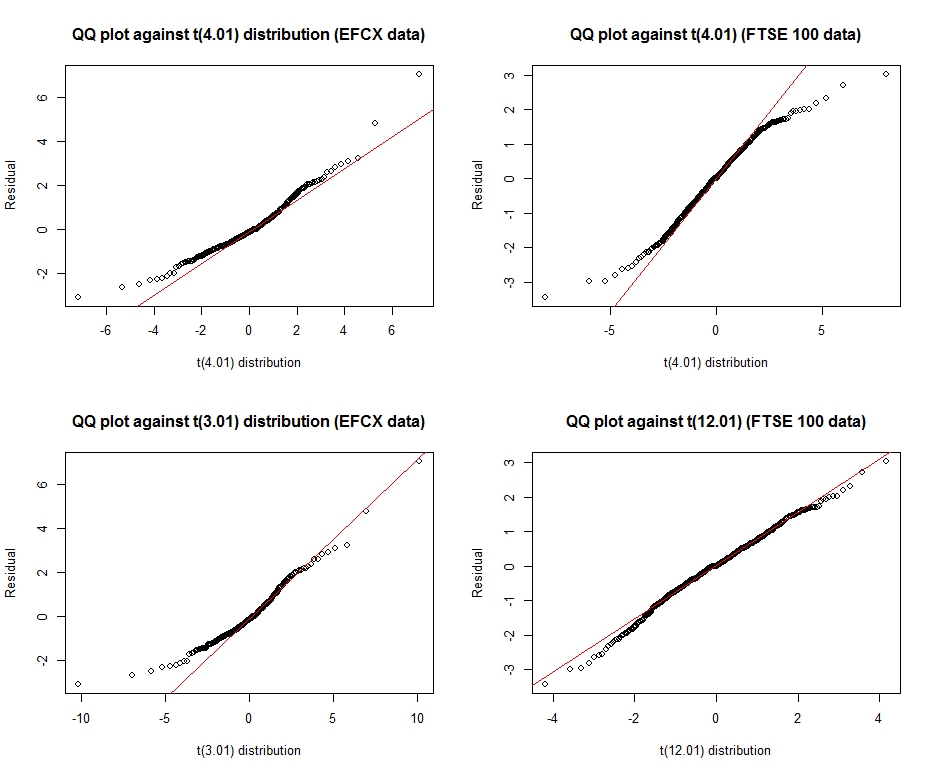}
\label{fig:QQplot}
\end{figure}

The top-left panel of Figure~\ref{fig:QQplot} shows the QQ-plot of the residuals against the $t(4.01)$ distribution for the EFCX data. The plot indicates a heavier-than-$t(4.01)$ upper tail, which implies that the fourth moment of the error term may not be finite. On the other hand, the QQ-plot against the $t(3.01)$ distribution (bottom-left panel of Figure~\ref{fig:QQplot})  yields a lighter-than-$t(3.01)$ lower tail---an indication  that ${\mathrm E}|\epsilon|^3 < \infty$.\vspace{-3mm}
                  
\vspace{-3mm}

For the FTSE 100 data, the QQ-plot against the $t(4.01)$ distribution in the top-right panel of Figure \ref{fig:QQplot} shows that the residuals may have lighter-than-$t(4.01)$ tails. The fit in the QQ-plot against the $t(12.01)$ distribution (bottom-right panel of Figure \ref{fig:QQplot}) looks quite good, from which we may  conclude that ${\mathrm E}|\epsilon|^4 < \infty$ holds for the FTSE 100 data. This might explains why all M-estimates of $\beta$ in Table \ref{FTSE.Mest} yield similar values.

\section{Conclusion}\label{sec.concludeMest}
We consider a class of M-estimators and the weighted bootstrap approximation of their distributions for the GARCH models. An iteratively re-weighted algorithm for computing the M-estimators and their bootstrap replicates are implemented. Both simulation and real data analysis demonstrate superior performance of the M-estimators for the GARCH~(1, 1), GARCH~(2, 1) and  GARCH~(1, 2) models. Under heavy-tailed error distributions, we show that the M-estimators are more robust than the routinely-applied QMLE. We also demonstrate through simulations that the M-estimators work well when the true GARCH~(1,~1) model is misspecified as the GARCH~(2, 1) model. Simulation results indicate that under the finite sample size, bootstrap approximation is better than the asymptotic normal approximation of the M-estimators.


\end{document}